\documentclass[fleqn,usenatbib]{mnras}

\usepackage[T1]{fontenc}

\DeclareRobustCommand{\VAN}[3]{#2}
\let\VANthebibliography\thebibliography
\def\thebibliography{\DeclareRobustCommand{\VAN}[3]{##3}\VANthebibliography}

\usepackage{graphicx}	
\usepackage{amsmath}	
\usepackage{amssymb}	
\usepackage{subcaption}
\usepackage{booktabs}
\usepackage{multirow}
\usepackage[dvipsnames]{xcolor}

\newcommand*{\kms}{\text{km}\,\text{s}\ensuremath{^{-1}}}
\newcommand*{\kcor}{\textit{K}\text{-correction}}
\newcommand*{\kcors}{\textit{K}\text{-corrections}}
\newcommand*{\kcored}{\textit{K}\text{-corrected}}

\newcommand*{\http}[1]{\href{http://#1}{#1}}

\usepackage{newtxtext,newtxmath}

\title[Empirically-Driven K-corrections]{Empirically-Driven Multiwavelength K-corrections At Low Redshift}

\author[C. E. Fielder et al.]{
Catherine E. Fielder,$^{1,2}$\thanks{E-mail: cfielder@arizona.edu}
Brett H. Andrews,$^{2,3}$
Jeffrey Newman,$^{2,3}$
Samir Salim $^{4}$
\\
$^{1}$Steward Observatory, University of Arizona, Tucson, AZ, 85721, USA\\
$^{2}$Department of Physics and Astronomy, University of Pittsburgh, Pittsburgh, PA 15260, USA\\
$^{3}$Pittsburgh Particle Physics, Astrophysics, and Cosmology Center (PITT PACC), University of Pittsburgh, Pittsburgh, PA 15260, USA\\
$^{4}$Department of Astronomy, Indiana University, Bloomington, IN, 47405, USA
}

\date{Accepted XXX. Received YYY; in original form ZZZ}

\pubyear{2022}

\begin{document}
\label{firstpage}
\pagerange{\pageref{firstpage}--\pageref{lastpage}}
\maketitle

\begin{abstract}
\kcors, conversions between flux in observed bands to flux in rest-frame bands, are critical for comparing galaxies at various redshifts. These corrections often rely on fits to empirical or theoretical spectral energy distribution (SED) templates of galaxies. However, the templates limit reliable \kcors\ to regimes where SED models are robust. For instance, the templates are not well-constrained in some bands (e.g., WISE W4), which results in ill-determined \kcors\ for these bands. We address this shortcoming by developing an empirically-driven approach to \kcors\ as a means to mitigate dependence on SED templates. We perform a polynomial fit for the \kcor\ as a function of a galaxy's rest-frame colour determined in well-constrained bands (e.g., $^{0}(g-r)$) and redshift, exploiting the fact that galaxy SEDs can be described as a one parameter family at low redshift ($0.01<z<0.09$). For bands well-constrained by SED templates, our empirically-driven \kcors\ are comparable to the SED fitting method of \textsc{Kcorrect} and SED template fitting employed in the GSWLC-M2 catalogue (the updated medium-deep GALEX–SDSS–WISE Legacy Catalogue). However, our method dramatically outperforms the available SED fitting \kcors\ for WISE W4. Our method also mitigates incorrect template assumptions and enforces the $\kcor$ to be $0$ at $z = 0$.  Our \kcored\ photometry and code are publicly available.

\end{abstract}




\section{Introduction}
\label{sec:intro}

Broad-band luminosity measurements are critical for understanding galaxy evolution, but require correcting for the redshifting of light across photometric band passes. The redshifting of a galaxy spectrum is equivalent to shifting the filter transmission curve through which the galaxy is observed. Hence two galaxies with identical rest-frame spectral energy distributions (SEDs) but at different redshifts will generally have different fluxes in the same observed band pass. This difference is referred to as the \kcor\ \citep{humason1956,oke1968}. \kcors\ are especially important for comparing populations of galaxies at varying redshift, as we want equivalent measurements of the SEDs of these galaxies at all $z$. \kcors\ serve as a critical equaliser, enabling practical comparisons amongst galaxy surveys at different redshift ranges. 

For objects where the entire spectrum has been observed, \kcors\ are straightforward to calculate, as the spectrum can be shifted to accommodate redshift corrections. However, the spectra obtained are frequently limited to the central region of a galaxy, so the appropriate \kcor\ for integrated photometry will likely differ. Since spectroscopy with full spatial coverage (or, better yet, spatially-resolved spectroscopy) is rarely available for large survey samples, we must depend on photometric information to derive \kcors.  In this paper, we develop new methods that can be used to derive \kcors\ based upon photometric measurements, even when we lack not only observed spectral measurements but even theoretical template spectra that may be used to guide predictions.

In order to determine the intrinsic brightness of an object in some rest-frame passband we use a transformation that acts upon the observed-frame photometry. The \kcor\ between a bandpass $R$ used to observe a galaxy at redshift $z$ and rest bandpass $Q$ may then be defined by 
 the equation \citep{oke1968,hogg2002,blanton2007}
\begin{equation}
    M_{Q} = m_{R} - DM(z) - K_{QR}(z) + 5\log{h},
\label{eq:kcor}
\end{equation}
where $M_{Q}$ is the absolute magnitude in the desired rest-frame passband $Q$, $m_{R}$ is the observed apparent magnitude in passband $R$, $DM(z)$ is the distance modulus derived assuming a Hubble parameter $H_0 = 100$ km/s/Mpc, $K_{QR}(z)$ is the \kcor\ between band $R$ and band $Q$, and $h$ is the adopted value of the Hubble parameter divided by 100 km/s/Mpc. The distance modulus is defined as $DM(z) = 5\log{(\frac{d_{L}}{10pc})}$ where $d_{L}$ is the luminosity distance. For brevity we do not include the full and rigorous formalism for the generalised definition of the \kcor, which is presented in \citet{hogg2002} (see in particular Equations 8 and 10).

The key ingredients generally used to determine \kcors\ for a given galaxy are (1) some knowledge of the spectrum of the object, which characterises the flux emitted by the galaxy across a large wavelength range, as well as (2) the response curves of the instrument used to make the observations, as the transmission function of different filters, the QE of different detectors, etc. will affect the collected photons slightly differently. Often we do not have much information on the full spectrum of the objects we observe; instead, we merely have a handful of photometric points that sample various parts of the observed spectrum. This requires us to either use analytical approximations to quantify how the observed galaxy's colour changed with redshift, or to utilise templates to reconstruct the observed galaxy's rest-frame SED. 

Historically, template matching has been the most commonly used approach. There are two families of SED templates that are used in this matching to determine \kcors : empirical templates derived from a representative set of real galaxy spectra \citep[e.g.,][]{coleman1980,kinney1996,brown2014}, or synthetic templates derived from stellar population synthesis (SPS) models \citep[e.g.,][]{bruzual2003,maraston2005}. SPS models employ assumptions about the star formation and evolution within a galaxy and libraries of stellar spectra to construct a realistic SED. As a result SPS templates have the benefit of providing estimates of other physical quantities such as stellar mass, star formation rate, or dust attenuation in addition to $K$-corrected photometry. 

There are a number of methods which have been used to calculate galaxy \kcors\ in the literature. We briefly review $\kcor$ methods here, but this list is not exhaustive. 
We then discuss the caveats to using these methods.

Early papers modelled \kcors\ as a simple function of redshift and galaxy morphological type, such as in the work by \citet{coleman1980}. The authors used empirical SED templates for four morphological galaxy types (E, Sbc, Scd and Irr galaxies) to obtain \kcors\, using the best data available at the time. However, this method is somewhat oversimplified based on our current understanding of galaxy evolution, as galaxies that may share the same morphology can have a diverse range of spectral properties \citep[see, e.g.,][]{uzeirbegovic2022}.

It is more common to obtain \kcors\ by modelling galaxy SEDs as a function of wavelength.   This process begins by identifying a template rest-frame SED that matches the observed colours of a galaxy when shifted to that galaxy's redshift.  The template may then be used to derive \kcors\ and determine the rest-frame properties of the galaxy of interest. Maximum likelihood fitting of photometry to theoretical SPS-based SED models has  often been utilised to calculate \kcors\, such as in work by e.g., \citet{bell2004} and \citet{brown2007}. As mentioned above, SPS models can provide reasonable approximations of real galaxy SEDs within the wavelength range of the spectral libraries used. However, while this method works very well for intrinsically red galaxies, the rest-frame spectra of blue galaxies are not as tightly constrained by broad-band photometry due to the effects of star formation, dust attenuation, and nebular emission lines which may have degenerate impacts on galaxy SEDs within the optical wavelength range but different impacts at shorter or longer $\lambda$. Even worse, existing SPS models may struggle to capture the full diversity of galaxy SEDs. \citet{blanton2003} introduced $\kcor$ calculations derived from matching photometric observations to templates constructed from \textit{combinations} of \citet{bruzual2003} SPS models, dust attenuation models, and nebular emission lines, which resulted in more realistic SEDs. This was the first instance of using multiple components to construct galaxy SEDs, which has served as the backbone of the commonly used \textsc{Kcorrect} software of \citet{blanton2007} (now in version 4.3). \textsc{Kcorrect} has become a widespread standard for determining \kcors\ and will serve as a comparison for our own results. 

\kcors\ have also been estimated using analytical functions of redshift which are parameterised in terms of some additional property that characterises the galaxy, separating out objects of different rest-frame SEDs. \citet{willmer2006} fit second-order polynomials to determine \kcors\ as a function of a single \textbf{observed} colour and redshift, based upon the \citet{kinney1996} empirical templates. This work was based upon photometry in only three bands (the CFHT 12K $BRI$ filters used for DEEP2, the Deep Evolutionary Exploratory Probe; \citealt{newman2013}) and used filters for which \textsc{Kcorrect} had not yet been tuned and tested (the fitting methods employed by \citet{blanton2003} were not available) resulting in the need to develop other methods. \citet{chilingarian2010} empirically showed that \kcors\ can be effectively approximated as solely a function of redshift and a single colour, using 190,275 observed galaxies spanning the redshift range $0.03<z<0.6$. The authors derived \kcors\ defined by polynomials that are fifth order in observed $(g-r)$ colour and third order in redshift for nine filters ($ugrizYJHK$), based upon both \textsc{Kcorrect} and \textsc{PEGASE}.2 \citep{fioc1997} SPS models. Work by \citet{omill2011} took a similar approach, deriving a linear relationship between the \kcor, $(g-r)$ colour, and redshift for a number of galaxies of SDSS DR7 that were processed by \textsc{Kcorrect v4.3}. \citet{beare2014} did a more in-depth study to determine which observed colours best characterise \kcors\ for each SDSS optical band ($ugriz$), deriving second order polynomials in colour from the 129 empirical SEDs in the \citet{brown2014} Atlas.


While \kcors\ often can be determined accurately using template fits, there are also limitations to this approach. First, analyses generally employ only a limited number of template SEDs (or linear combinations thereof). As a result, the templates used may not span the full range of possible SEDs exhibited by a diverse galaxy sample. This is especially concerning for galaxies that have less commonplace SEDs, which could result in inaccurate \kcors\ due to template mismatches. Second, although empirical templates may provide a better match to the SEDs of actual galaxies than model-based ones, the existing empirical template libraries only have limited wavelength coverage, particularly in the infrared, limiting their use. Additionally, many of the older commonly used empirical templates such as those from \citet{coleman1980} and \citet{kinney1996} are derived from galaxies with significant light from active galactic nuclei (AGN). This can lead to systematic errors due to the strong emission lines and blue continuum from the AGN outshining the stellar continuum. This problem is exacerbated for empirical templates that are based on the spectra of the  centres of bright galaxies. 
 \citet{willmer2006} found that the use of such templates (for instance, several of the \citealt{kinney1996} templates) resulted in derived colours that are too red by $0.08$ magnitudes.  Empirical templates are also still limited by the small sets of galaxies for which robust observations over a broad wavelength range have been obtained, meaning some galaxy families do not have well-observed template SEDs. Theoretical SEDs may also cover a limited wavelength range for the same reason. 
Methods that employ model-based templates are also limited by the quality of the models used. 

Regardless of which family of template is used, depending on them restricts our ability to determine \kcors\ at wavelengths where the templates are not available, poorly constrained, or models are lacking. For example, the WISE bands probe the interstellar dust content of galaxies due to their sensitivity to polycyclic aromatic hydrocarbon emission features, the small-grain dust continuum, and the thermal emission tail of larger dust grains \citep{wright2010}, which makes them important for studying a galaxy's characteristics. There are currently few options available for determining WISE \kcors\ apart from \textsc{Kcorrect}. However, while \textsc{Kcorrect} will calculate \kcors\ for the WISE photometric bands, these fits employ templates that do \textit{not} include dust emission features, which are increasingly important for the longer-wavelength WISE bands ($W2$ through $W4$). The absence of this feature in the templates should result in incorrect \kcors . 


In this work we have developed a data-driven approach to determining \kcors\ that limits the extent to which templates are relied upon and that circumvents entirely the need for templates in filters where the SED is poorly known. Conceptually, our method builds on the work of \citet{willmer2006}, \citet{omill2011} and \citet{brown2014}, which modelled the \kcor\ needed for a given band as a simple (polynomial) function of a galaxy's redshift and a quantity that characterises the SED of the galaxy. However, in this work we use the rest-frame colour for some pair of bands as that parameter, rather than an observed colour. This rest-frame colour can be determined using templates in a spectral regime where the range of galaxy SEDs is very well constrained, such as the optical rest-frame $(g-r)$ band. 

If we make the assumption that at low redshift galaxy SEDs can be approximated into a single parameter family -- equivalent to assumptions made in that previous work --  we show that with a single rest-frame colour we can match objects with equivalent SEDs across redshift, and thereby derive the dependence of $K$-corrections on the chosen rest-frame colour and redshift. With this approach, so long as one has access to a rest-frame colour that has been $K$-corrected already, one can then determine rest-frame colours for any other bands of interest. We caution that this assumption has only been tested extensively at low redshift ($z<<1$) with a limited $\Delta z$ range; the situation may become more complex at higher redshift, where the assumption that SEDs fall into a single parameter family \citep{connolly1995,madgwick2003a,madgwick2003b} or that \kcors\ can be approximated as low-degree polynomials may break down. We presented an overview of this approach in \citet{fielder2021}, where we employed it to determine rest-frame WISE photometry; we describe and test this method more extensively in this work. 

This paper is organised as follows. In \autoref{subsection:data}, we briefly described the data used in our calculations, which includes GALEX, SDSS, 2MASS, and WISE observations. \autoref{subsection:calculation} details our method for determining \kcors. In \autoref{sec:results} we compare our computed rest-frame colours from our \kcors\ to literature values. Finally, we summarise and discuss our findings in \autoref{sec:conclusions}.

All magnitudes and colours are presented in the AB system in this work. Absolute magnitudes are determined using a Hubble constant $H_{0} = 100 \;\kms \rm{Mpc}^{-1}$, so they are equivalent to $M_{y} - 5\log{h}$ (where $M_{y}$ is the $y$-band absolute magnitude and $h = H_{0}/100$) for other values of $h$. Photometry is presented adopting the notation used in \citet{blanton2007} and \citet{licquia2015b}, where an absolute magnitude corresponding to passband $y$ as observed at redshift $z$ is denoted as $^{z}M_{y}$. 


\section{Methods}
\label{sec:methods}

\subsection{Observational Data}
\label{subsection:data}

The data used in this study originates from the GALEX-SDSS-WISE-Legacy Catalogue 2 (GSWLC-2) of \citet{salim2016,salim2018}, within the redshift range of $0.01 < z < 0.09$. This data spans across the filter range of the GALEX ultraviolet survey $FUV/NUV$ bands \citep{galex2005}, the SDSS optical $ugriz$ bands (DR10; \citealt{ahn2014}), the 2MASS near-infrared $JHKs$ bands \citep{skrutskie2006}, and the WISE mid-infrared $W1/W2/W3/W4$ bands \citep{wright2010}. Specifically, the WISE data is from the ``unWISE'' reduction of \citet{lang2016} which is more appropriate for galaxies than the official pipeline. We use photometric data from the medium-deep GSWLC catalogue, where the depth pertains to the UV photometry from GALEX, which unlike other bands is generally of much less uniform depth - going from shallow to very deep. While using the medium-deep instead of the shallow catalogue decreases the number of galaxies at our disposal by roughly one half, the increased signal-to-noise in the UV imagining is a worthwhile trade-off. 

For objects within the GSWLC-M2 catalogue we have both the observed photometry (and associated errors) in addition to rest-frame values from the UV to the near IR (excluding WISE). These rest-frame values were obtained via fits to SED models, as described in \citet{salim2016}, which we summarise in \autoref{subsection:kcorrection_catalogues}. For convenience we convert the observed flux ($f_{\nu}$) in Jy to AB magnitudes using the relation $m_{\rm{AB}}=-2.5\log_{10}{(f_{\nu})} + 8.90$. As a result all magnitudes presented in this work are in the AB system. 

In order to keep our galaxy sample at a maximum we split our data into four different catalogues: (1) a catalogue consisting of the SDSS $ugriz$ optical bands, (2) a catalogue consisting of the GALEX $FUV/NUV$ bands and the $ugriz$ optical bands which we call UV+optical, (3) a catalogue consisting of the 2MASS $J/H/Ks$ near-IR bands and the $ugriz$ optical bands which we call near-IR+optical, and (4) a catalogue consisting of the WISE $W1/W2/W3/W4$ mid-IR bands, the $J/H/Ks$ near-IR bands, and the $ugriz$ optical bands which we call IR+optical. This way we can ensure that results from runs of \textsc{Kcorrect} are not driven by a specific set of bands. For example, an IR prediction can be entirely driven by a fit in the UV+optical depending on the signal-to-noise of the measurement, particularly because extrapolation is common for IR SED models.

Before we take further steps in our analysis we exclude all objects that have photometric values of ``NaN'', infinity, or $0$, which all indicate missing photometry. When evaluating the fits for the \kcors\ we use a limited redshift range $0.04<z<0.09$ in order to avoid selection effects at low redshift ($0.01<z<0.04$), but compute \kcors\ for the full redshift range of our sample. For the fits in the $FUV/NUV$ bands we also exclude objects that have large GALEX photometric errors. Specifically, in $FUV$ we perform the \kcor\ fits restricting to galaxies with errors $\sigma_{m_{FUV}}<0.12$, while in $NUV$ we require errors $\sigma_{m_{NUV}}<0.1$. We apply similar cuts for objects that have large WISE photometric errors. These correspond to $\sigma_{m_{W1}}<0.013$, $\sigma_{m_{W2}}<0.025$, $\sigma_{m_{W3}}<0.15$, and $\sigma_{m_{W4}}<0.25$. These cuts were determined by examining distributions of the errors and how they biased the \kcor\ results. Before analysing our results we do a final clean to remove any ``NaN'' or infinity from each catalogue for the rare cases that an object did not have a proper fit in \textsc{Kcorrect} or our own analysis. Within our final wavelength separated catalogues the optical catalogue contains 148,704 galaxies, the UV+optical catalogue contains 28,318 galaxies, the near-IR+optical catalogue contains 74,038 galaxies, and the IR+optical catalogue contains 40,361 galaxies. Each spans from redshift $0.01 < z < 0.09$. The catalogues containing our $K$-corrected results and comparison values are publicly available at \href{https://github.com/cfielder/K-corrections/tree/main/Catalogs}{the catalogue section of our catalogue GitHub page.}\footnote{https://github.com/cfielder/K-corrections/tree/main/Catalogs}.

\subsection{\textit{K}-corrections from Previous Work}
\label{subsection:kcorrection_catalogues}

We will be comparing our results from our fit calculations (described in the following subsection \autoref{subsection:calculation}) to \kcors\ from two other sources as a means to validate our results. 
\begin{enumerate}

    \item We generated \kcors\ with the \textsc{kcorrect v4.3} software package \citet{blanton2007}. \textsc{Kcorrect} is a robust method for determining \kcors\ using a combination of SED templates with SPS models, dust models, and nebular emission models. \textsc{Kcorrect} uses a non-negative matrix factorisation algorithm that creates model based template sets. A set of five template SEDs are used, which are derived from combinations of 450 \citet{bruzual2003} SPS models across a wide range of age and metallicity, and 35 ionised gas emission models from \citet{kewley2001}. After the templates are generated and reduced into 5 sets, linear combinations of these templates are fit to the measured photometry and respective photometric errors for each galaxy for which \kcors\ are being calculated with a $\chi^{2}$ minimisation technique. This constructs an estimated SED for that galaxy. From the estimated SED \kcors\ and other physical parameters can be determined. 
    
    We ran \textsc{Kcorrect} on the observed photometry and associated errors within the GSWLC-M2 catalogue for bands from the UV to IR composed of combinations of subsets of the wavelength range as described in \autoref{subsection:data} in addition to the full wavelength range. \textsc{Kcorrect} can compute \kcors\ for all filters considered in this paper ($FUV$ through $W4$). We added an additional error term in quadrature to our inverse variance maggies equivalent to $1\%$ of the respective observed maggy for the given band to allow for more flexibility in the models. 

    \item GSWLC \citep{salim2016,salim2018} produces \kcors\ by performing SED fitting using model SEDs. For this catalogue a Bayesian approach to SED fitting was employed (\citealt{salim2007}, see also \citealt{kauffmann2003,brinchmann2004,tremonti2004} for the basis of this method) utilising the CIGALE SED fitting code \citep{boquien2019}, which used a combination of \citep{bruzual2003} SPS models, dust models, and nebular emission models. In this approach millions of distinct models are considered individually, rather than utilising a linear combination of a smaller number of models (templates) like that performed by \textsc{Kcorrect}. Additionally, the GSWLC fits employ more recently updated models. In contrast to the best-fit ($\chi^{2}$ minimisation) approach of SED fitting, such as that employed by \textsc{Kcorrect}, the Bayesian approach determines the full probability distribution of any parameter. This has the advantage of more robust parameter characterisation and uncertainty. The absolute magnitudes derived in the GSWLC catalogues are determined from the best fit model by comparison between the redshift model fluxes and the observations.
    
    We simply adopt these rest-frame magnitudes directly from the catalogue, as the GSWLC-M2 catalogue serves as the basis for the data used in the analysis presented here so no additional conversions or calculations are necessary.

\end{enumerate}

Results from \textsc{Kcorrect} and GSWLC-M2 provide two standards of comparison for our own derived \kcors. One of the innate difficulties in determining \kcors\ is the lack of a "true" answer for any observed galaxy. This makes it challenging to probe any differences between our own data driven approach and \kcors\ from other template based approaches. Our goal is a method for which results are comparable to those from other approaches in regimes where we trust the SED templates and a technique that is easier to implement than those used by \textsc{Kcorrect} and GSWLC.

\subsection{Deriving Data-Driven \kcors}
\label{subsection:calculation}

Our approach to determining \kcors\ is based on the simplifying assumption that galaxy SEDs at low redshift can be approximated as a single-parameter family. This assumption stems from the results of applications of dimensionality reduction methods such as principal component analysis, which have found that the range of galaxy spectra at low redshift can largely be described with only a single parameter \citep{connolly1995,madgwick2003a, madgwick2003b}. Likewise, previous work has shown that \kcors\ can be approximated by an analytical function of redshift parameterised by a single quantity that characterises the intrinsic galaxy SED such as an observed colour \citep{chilingarian2010, omill2011, beare2014} or $D_{n}4000$ \citep{westra2010}. 

In this work, we leverage rest-frame colour information as the single parameter of choice; i.e., it serves as our means of distinguishing objects with intrinsically different SEDs.  By choosing a rest-frame colour that can be well-determined by other \kcor\ methods, we can then bootstrap off of that reference colour to determine \kcors\ for the rest of the SED. While the assumption of this direct mapping of a single rest-frame colour to SED shape is not perfect due to the ambiguity of contributions to the SED at fixed rest-frame colour \citep[e.g.,][]{conroy2013,magris2015}, it is sufficient to first order out to $z=1$ \citep{madgwick2003b}; in our primary applications, we consider galaxy samples that only span from $0.01 < z < 0.09$, so this is a reliable assumption.

Our approach for calculating \kcors\ is to construct a function that produces rest-frame absolute magnitude given (1) the apparent magnitude of the band that needs to be corrected; (2) the apparent magnitude in a second band that serves as an ``anchor''; (3) rest-frame colour in some pair of bands (which may include the anchor band but not the band which needs correcting; this rest-frame colour may be determined via any \kcor\ method of choice); and (4) redshift. \kcors\ for the anchor band are also needed if we wish to determine rest-frame absolute magnitudes for the band that we wish to correct, but are not necessary if we only wish to determine rest-frame colours or SED shapes. 
We opt to use high signal-to-noise bands for the reference rest-frame colour and the anchor band; hence,  we generally employ $(g-r)$ and $m_{r}$ in this work. Some caution is warranted in this choice, however, as galaxies that span a large range in $(NUV-r)$ (4--6; i.e., green valley and truly quiescent galaxies) can end up with similar $^{0}(g-r)$ values around $\sim0.75$. However, for the UV bands we have tested other choices for rest-frame colours such as $^{0}(u-r)$ and found comparable \kcored\ results to when $^{0}(g-r)$ is employed.

We approximate our \kcors\ as a polynomial function of redshift with coefficients that are dependent upon rest-frame colour, multiplied by the observed (anchor - target band) colour. Thus, we utilise \kcors\ that are defined by:
\begin{equation}
   K_{\rm{corr}} = f(z, ^{0}(A-B)),
\end{equation}
where $z$ is redshift and $^{0}(A-B)$ is a known rest-frame colour, $f(z, ^{0}(A-B))$ is a polynomial function of redshift to be determined.

To determine the parameters of the function $f(z, ^{0}(A-B))$ used to derive rest-frame colours, we perform a series of polynomial fits that are outlined in the following algorithm. These fits are tempered by various error cuts that determine which quantities are used for which fits, as described below. In the following, $Y$ will refer to our generalised target band, which can be any band of interest except $r$, which we assume to be the anchor band.

\begin{enumerate}
    \item \textit{The \textbf{initial fits}: Determining the dependence of the observed (anchor - target band) colour on redshift, in bins of rest-frame colour.}\\  
    
    We first wish to determine the polynomial relationships between the \textit{observed} colour in a pair of bands (one of which serves as the anchor, and one of which is the desired target band) and redshift, for bins of fixed \textit{rest-frame} colour in another (e.g., optical) pair of bands. The relationship between these three quantities is exemplified in \autoref{fig:color_vs_z}, where we plot observed $(r-i)$ colour as a function of redshift, with the points coloured according to each galaxy's rest-frame $(g-r)$. It is clear that the change in $(r-i)_{\rm obs}$ with $z$ depends upon both rest-frame colour and redshift. We analyse this dependence using small bins of rest-frame colour ($\sim20$ in total) with an approximately equal number of objects per bin. In the work presented here we use bins of rest-frame $(g-r)$ colour, but any other reliable colour measurement can be used. 
    
    In each bin we perform a polynomial fit to determine the relationship between observed colour and redshift at fixed rest-frame colour. We compute both linear and quadratic fits for all bands for comparison purposes but note that our work favours the linear approach in the cases we have explored to date, as discussed in \autoref{subsection:extrapolation}. Our plots and tables generally provide results from linear fits for the initial polynomials only. The function we fit in each bin is defined as
    \begin{equation}
        (r - Y)_{\rm obs} = f(z) = a_{0} + a_{1}z + a_{2}z^{2},
    \label{eq:polynomial}
    \end{equation}
    where $a_{0}$ is the constant term, $a_{1}$ is the linear coefficient, and $a_{2}$ is the quadratic coefficient. We fit for the parameters of \autoref{eq:polynomial} using a Huber regression technique \citep{huber1964}, which is robust to outliers. These fits are implemented via $\tt{scikit-learn}$ ($\tt sklearn.linear\_model.HuberRegressor$; \citealt{scikit-learn}), where we set the parameter that controls the number of samples to be counted as outliers $\epsilon=1.01$. A smaller $\epsilon$ will provide results that are more robust to outliers (one is the minimum value). For the linear case $a_{2}$ is set to 0. Thus, for each bin in rest-frame colour we determine the fit coefficients $a_{0}$, $a_{1}$, and $a_{2}$ simultaneously (or just $a_{0}$ and $a_{1}$ for the linear case). 

    Although we leave $a_0$ free when fitting, in the remainder of our procedures we fix its value at zero. This is necessary because when $f(z)$ is evaluated at a redshift of zero, its value would correspond to $a_{0}$.  As we have defined it, $f(z)$ corresponds to the offset between observed colour and rest-frame colour. However, \textbf{at redshift zero the \kcors\ for all bands themselves must be zero}, as the observed and rest-frame bandpasses are identical in that case. Expressed in terms of \autoref{eq:polynomial}, it follows that $f(0) \equiv a_{0} \equiv 0$.
    
    In order to prevent higher-redshift objects from dominating the fits for observed colour vs. $z$ in the redder $^{0}(g-r)$ bins due to their greater numbers, we employ a redshift-dependent weighting. 
    To determine these weights, we split the objects within a single bin in $^{0}(g-r)$ into 40 bins in redshift space. Using these redshift bins and the counts in each bin $N_{bin}$ we build an interpolating function $N_{bin} = g(z_{cen})$, where $z_{cen}$ is the centre of the redshift bin, using the \citet{SciPy2020} $\tt{scipy.interpolate.interp1d}$ algorithm. With this function we can pass in the redshift of each individual galaxy to obtain the value of $N_{bin}$ at its $z$. 
    The weights we use are then defined as:
    \begin{equation}
        w = \frac{1}{(NMAD(r-Y)_{\rm{obs}})^{2}} \frac{<N_{bin}>}{N_{bin}},
    \end{equation}
    where $(NMAD(r-Y)_{\rm{obs}})$ is the normalised median absolute deviation (NMAD) of the observed colour and $<N_{bin}>$ is the mean number of galaxies across the $40$ bins. This down-weights bins with large numbers of objects, which has the greatest impact in the redder $^{0}(g-r)$ bins that are dominated by higher-redshift objects that otherwise skew the initial fits. These weights are then utilised in the aforementioned Huber regression of \autoref{eq:polynomial} by assigning them as $\tt{sample\_weight}$ $= w$.

    As an example of the initial fit results we show the linear fits for observed $(r-i)$ as a function of redshift in \autoref{fig:first_fit}. We have elected to present 3 bins in $^{0}(g-r)$ instead of the full set of 22 for clarity. Each bin contains $\sim 5435$ galaxies. The points show all of the galaxies within the respective bin, while the black dashed lines show the result of the linear fit using \autoref{eq:polynomial} ($a_{2}=0$). 
    While the bin with blue points could also be fit by a quadratic, overall \kcor\ results proved to be more stable when we restrict all fits to be linear; this may not prove true in future applications.  
    \\

    \item \textit{The \textbf{secondary fits}: Determining dependence of the initial fit coefficients upon rest-frame colour.}\\
    
    Having obtained initial fits for each bin in rest-frame colour, we next fit for the linear dependence of each of the necessary fit coefficients ($a_{1}$ and possibly $a_{2}$) on the \textit{mean} rest-frame colour within a bin.   Since the value of $a_{0}$ must be zero (as described above), we make small adjustments to the observed photometry based only upon $a_{1}$ (and, if necessary, $a_{2}$) to obtain $^{0}(r-Y) = (r-Y)_{\rm obs} + f(z)$.
    
    We hence only need to determine $a_{1}$ and (if relevant) $a_{2}$ as functions of rest-frame colour. We express these fits as 
    \begin{equation}
        a_{1} = b_{1}\langle^{0}(g-r)\rangle + b_{0}  
    \label{eq:a1_func}
    \end{equation}
    and
    \begin{equation}
        a_{2} = c_{1}\langle^{0}(g-r)\rangle + c_{0},
    \label{eq:a2_func}
    \end{equation}
    where $b_{0},\ b_{1},\ c_{0},$ and $c_{1}$ are fit coefficients and intercepts. 
    Therefore we assume a linear dependence between each coefficient and the rest-frame reference colour.  These fits are calculated using a Huber regression as before.  For these secondary fits we also tested polynomials up to fifth order as well as logarithmic and exponential functions, but found that a linear fit performed just as well if not better than these more complex functions.
    
    We exclude those colour bins for which the $a_{1}$ and/or $a_{2}$ values are highly uncertain from the fits to prevent them from influencing our results, as we found that in some bands the Huber regression produced unstable results when such bins were included.  For each $^{0}(g-r)$ bin we calculate the NMAD of the residuals from the initial fit (i.e., the NMAD of the differences between the  predicted $(r-Y_{\rm{obs}})$ value for each galaxy and its actual $(r-Y_{\rm{obs}})$). Bins where the NMAD was more than 2.5$\times$ larger than the minimum NMAD across all bins are excluded from the fitting of $a_{1}$ and $a_{2}$ as a function of mean rest-frame colour.  

    \autoref{fig:a1} illustrates the secondary fit for the $i$ band. We plot the $a_{1}$ values determined from the initial fits (from step (i)) as a function of the mean $^0(g-r)$ for each bin as blue points. The black dashed line shows the linear function resulting from Huber regression for this "secondary" fit (\autoref{eq:a1_func}). For this band, none of the $a_{1}$ values were excluded by the NMAD cutoff. 
    
    Once this fit has been completed, the \kcor\ for individual galaxies can be determined by calculating $a_1$ and $a_2$ based on each one's rest-frame colours and then evaluating the resulting function $f(z) = a_1 z + z_2 z^2$ using that object's redshift.
   \\

\begin{figure}
    \centering
    \includegraphics[width=\linewidth]{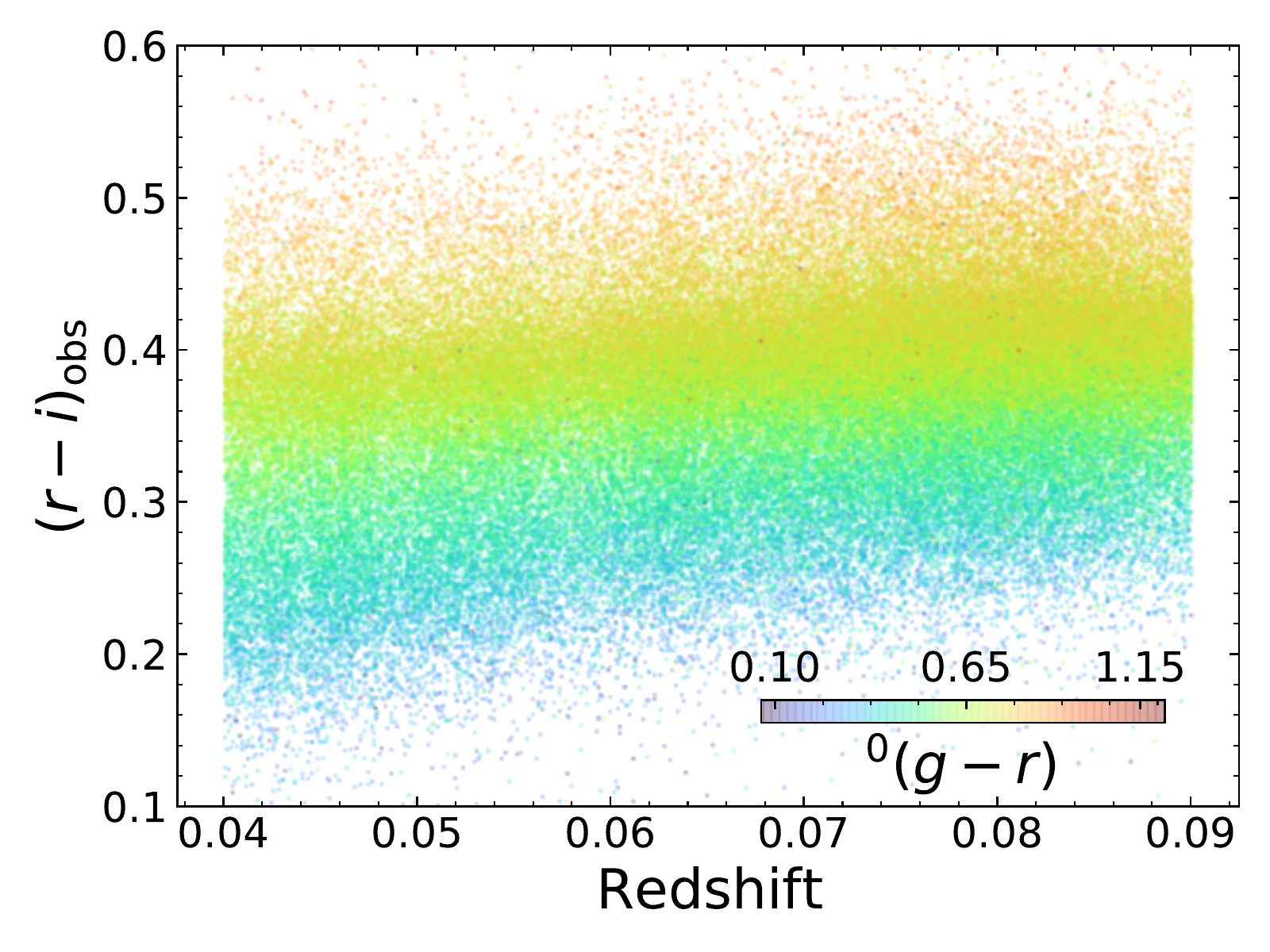}
    \caption{Observed $(r-i)$ colour plotted as a function of redshift for our galaxy sample. We have colour coded the points by rest-frame $^{0}(g-r)$ colour, where purple corresponds to more blue galaxies and red corresponds to more red galaxies. It is clear that observed colour is correlated with rest-frame colour and redshift. By determining fit coefficients in bins of rest-frame colour (cf. \autoref{fig:first_fit}) we can quantify how the observed colour from a given pair of bands changes relative to the rest-frame colour across redshift space. %
    }
    \label{fig:color_vs_z}
\end{figure}

\begin{figure}
    \centering
    \includegraphics[width=\linewidth]{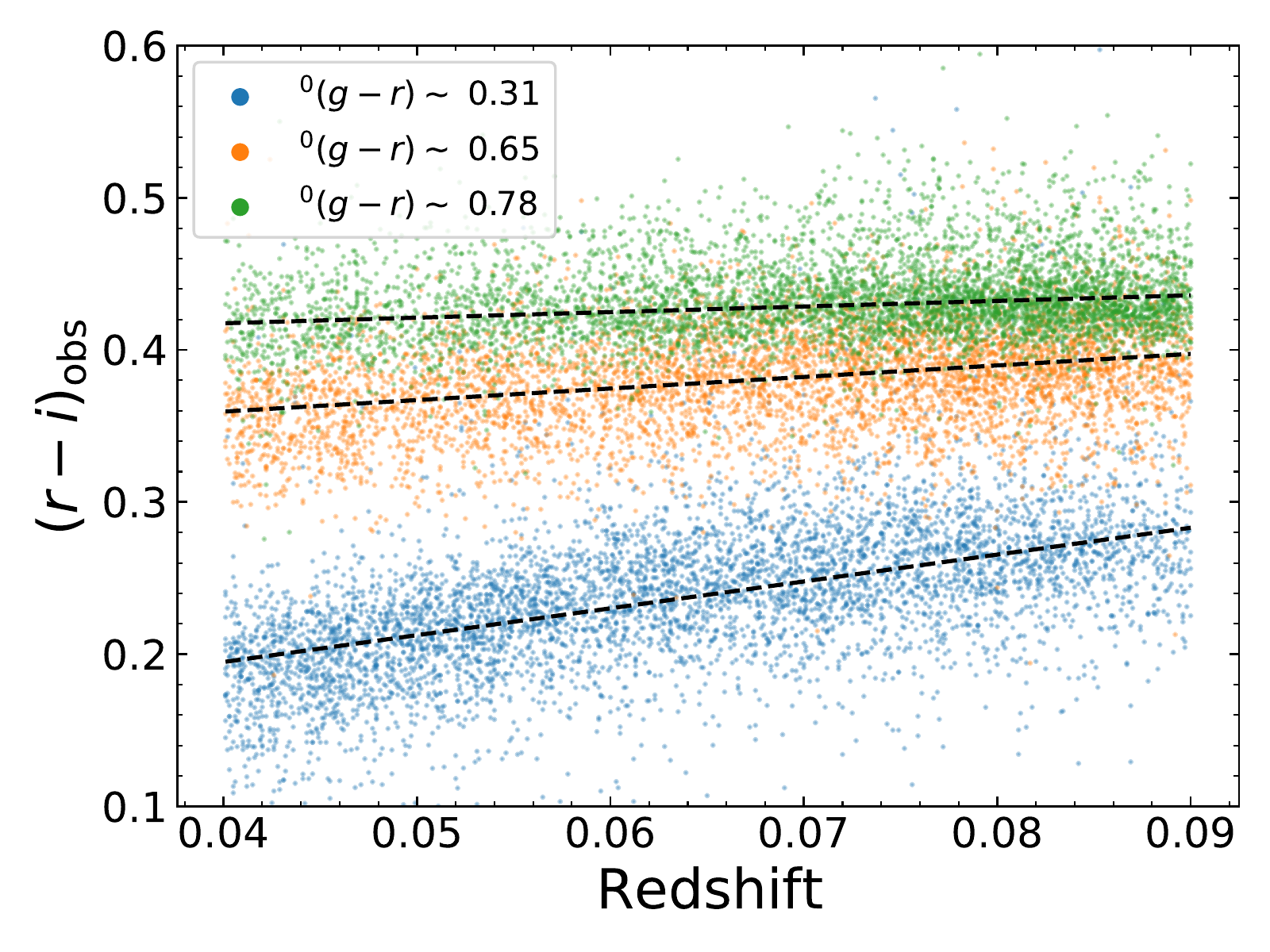}
    \caption{An example of our first-order polynomial fits to the redshift dependence of observed colour at fixed rest-frame colour. In this example we desire to $K$-correct $i$-band photometry, while the $r$-band serves as our anchor. We plot the observed $(r-i)$ colour as a function of redshift for three out of 22 equal-number $^{0}(g-r)$ bins constructed from the full sample; each bin includes $\sim5435$ objects. The bins whose data is plotted are labelled according to the average $^{0}(g-r)$ colour within each. The black dashed lines represent the fits to \autoref{eq:polynomial} for each bin plotted, performed using a robust Huber regression. The fits provide a reasonable representation of the trends within the data and are resistant to being pulled by outliers. %
    }
    \label{fig:first_fit}
\end{figure}

\begin{figure}
    \centering
    \includegraphics[width=\linewidth]{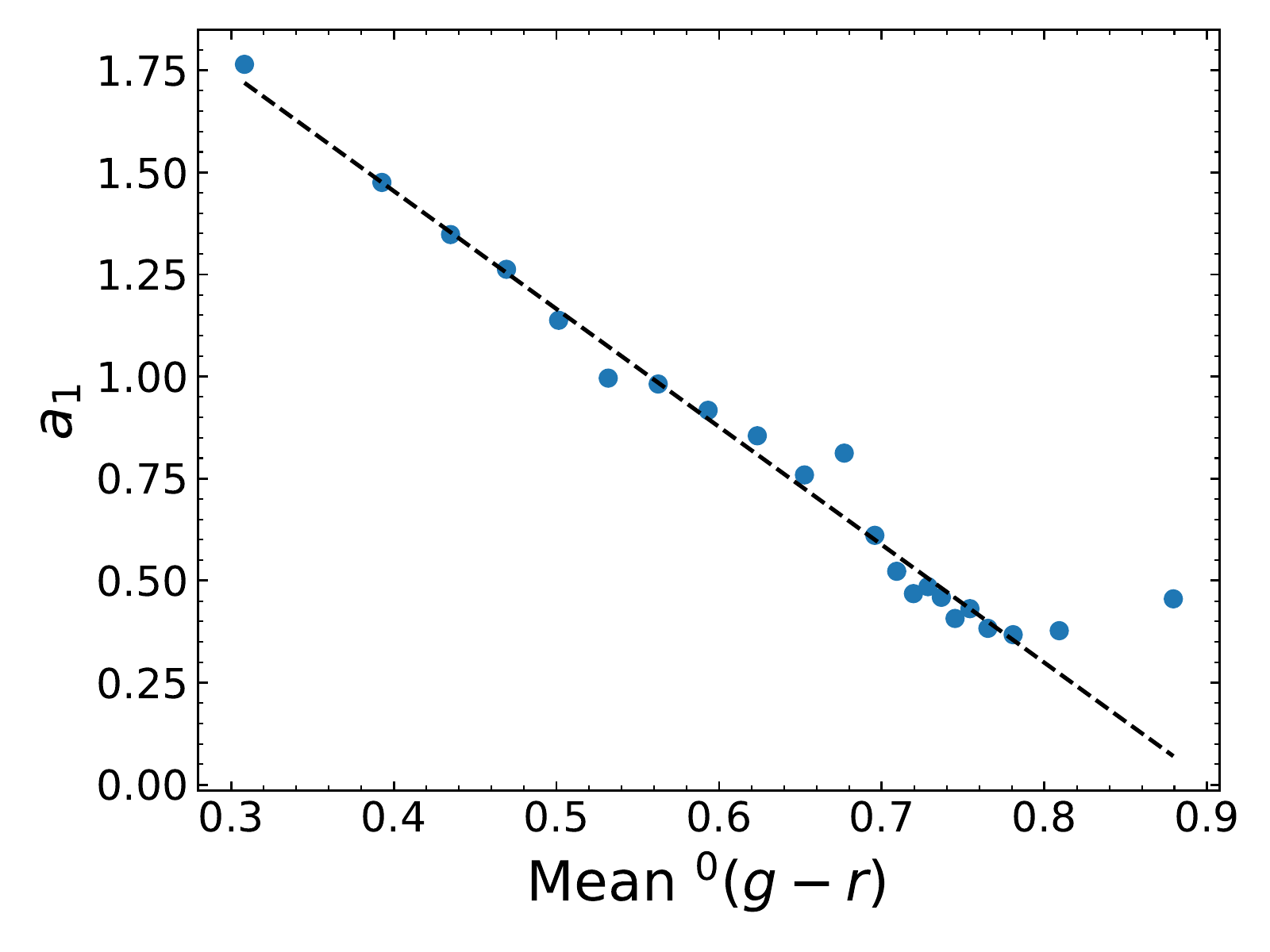}
    \caption{Linear-term coefficients ($a_{1}$) from the initial fit (i.e., the fit of observed colour as a function of redshift in bins of rest-frame colour, $^{0}(g-r)$, examples of which are shown in \autoref{fig:first_fit})  are plotted as a function of the central $^{0}(g-r)$ of each bin using blue points. 
    We perform a linear fit for $a_{1}$ as a function of the centre of a $^{0}(g-r)$ bins using a Huber regression (\autoref{eq:a1_func}), the result of which is plotted here as a black dashed line. A linear fit provides a good approximation to the plotted points. None of the $a_{1}$ values in this band were excluded from the fit by the NMAD cutoff, but the Huber regression is robust to the reddest and most discrepant point.
    }
    \label{fig:a1}
\end{figure}
    
    \item \textit{Determine whether to use a constant $a_{1}$ value instead of a linear fit.}\\
    As mentioned above, although we tested using but linear and quadratic initial fits, the final derived \kcors\ in this paper (as used in \autoref{fig:sed}) will \textit{all} be determined from purely linear fits (corresponding to $a_{2} = 0$ in \autoref{eq:polynomial}). However, even with that simplification, for some bands we obtain more stable results by using a constant value of the $a_{1}$ parameter instead of having $a_{1}$ depend linearly upon rest-frame colour as in \autoref{eq:a1_func}. In these cases the $a_{1}$'s resulting from the initial fit varied little from each other, causing the secondary fit to be approximately a flat line. For such bands, instead of calculating $a_{1}$ as a function of mean rest-frame colour as described in step (ii), we simply use the median of the set of $a_{1}$ values across all rest-frame colour bins. 
    
    To determine whether to use a constant $a_{1}$ or the results of the secondary fit we take an information criterion-based approach. First we determine the NMAD of the $a_{1}$ values resulting from the initial fits (obtained from \autoref{eq:polynomial}). We label this normalised median absolute deviation as $\sigma_{c}$; so $\sigma_{c} \equiv$ NMAD($a_{1}$). Then we determine the NMAD of the set of residuals $a_{1}-a_{1,\rm{pred}}$, which we label as $\sigma$. $a_{1,\rm{pred}}$ are the $a_{1}$ values predicted by the secondary fit for $a_{1}$ as a function of $^{0}(g-r)$ (obtained from \autoref{eq:a1_func}) for each $^{0}(g-r)$ bin. If the residuals are entirely due to Gaussian random errors, this statistic should approximate the standard deviation of the error distribution.
    
    
    The $\Delta\chi^{2}$ between the residuals for a constant $a_0$ versus those obtained when using the linear fit can then be approximated by
    \begin{equation}
        \Delta\chi^{2} = N(1 - \frac{\sigma_{c}^{2}}{\sigma^{2}}),
    \end{equation}
    where $N$ is the total number of $^{0}(g-r)$ bins used in the initial fit in step (i) that pass the NMAD cutoff in step (ii). Then we define the quantity $\Delta$AIC (which is based upon the Akaike Information Criterion, a quantity that can be used to test whether additional parameters meaningfully improve fits; \citet{akaike1974}) as
    \begin{equation}
        \Delta\rm{AIC} = 2 + \Delta\chi^{2},
    \end{equation}
    as the linear fit has a total of two free parameters. For photometric bands where $\Delta$AIC $<-10$ we use a linear fit for $a_{1}$; however, when $\Delta$AIC $\geq-10$, indicating that there is at most moderate evidence that a linear fit performs better, we instead set the $a_{1}$ value to a constant corresponding to the median of the $a_{1}$'s determined in step (i). 
    
    In this work, a constant $a_{1}$ value was favoured for the $J$, $H$,$Ks$, $W1$, $W2$, $W3$, and $W4$ bands when fits were performed for $a_1$ as a function of \textsc{Kcorrect} $^{0}(g-r)$ colour.  When instead GSWLC-M2 $^{0}(g-r)$ colour was used,  a constant $a_{1}$ was favoured for $z$, $Ks$, $W1$, $W2$, and $W4$. All other bands use what we refer to as a ``linear'' $a_{1}$ dependence on rest-frame colour, corresponding to \autoref{eq:a1_func}. \\

    \item \textit{Determine rest-frame colours and/or absolute magnitudes using the coefficient fits.}\\
    
    At this stage we now have a function that predicts the difference between the observed colour $(r - Y)$ and rest-frame $^{0}(r - Y)$ as a function of both an object's rest-frame colour in another pair of bands and the redshift. Since the value of $a_0$ must be zero (as described in step (i) above), we can re-write \autoref{eq:polynomial} as:
    \begin{equation}
        (r-Y)_{\rm{obs}} =\ ^{0}(r-Y) + a_{1}z + a_{2}z^{2}.
    \label{eq:polynomial2}
    \end{equation}
    
    Thus by subtracting $a_{1}z$ and $a_{2}z^{2}$ from the observed colour, the rest-frame colour in that pair of bands can be obtained; i.e., we have calculated a \kcored\ colour. We can solve for the rest-frame colour, obtaining: 
    \begin{equation}
        ^{0}(r-Y) = (r-Y)_{\rm{obs}}-a_{1}z-a_{2}z^{2} = m_{r} - m_{Y} - a_{1}z - a_{2}z^{2}.
    \label{eq:rest_color}
    \end{equation}
    
    If one is interested in determining the absolute magnitude in rest-frame band $Y$ instead of the rest-frame $r-Y$ colour, then one \textbf{must} have access to a \kcored\ version of the anchor band. To solve for $^{0}M_{Y}$ instead of $^{0}(r-Y)$, we exploit the fact that $^{0}(r-Y) = ^{0}M_{r} - ^{0}M_{Y}$, which we can rearrange to obtain:
    \begin{equation}
        ^{0}M_{Y} = ^{0}M_{r} + ^{0}(r-Y),
        \label{eq:absmag}
    \end{equation}
    where $^{0}(r-Y)$ can be determined from \autoref{eq:rest_color}.  $^{0}M_{Y}$ should then correspond to the \kcored\ absolute magnitude in band Y.\\
    
    \item \textit{Determine errors on \kcors\.}\\
    Having determined the values of the \kcors\ of interest, we next wish to estimate their uncertainties. Since the  results presented in this paper are all derived from either a linear fit for the $a_1$'s or by using a constant $a_{1}$ value, we describe our procedures for the case of linear fits; the methods we use can also be applied in other scenarios, however. 
    
    Specifically, we estimate the errors in \kcored\ photometry by applying propagation of errors to \autoref{eq:rest_color}. The resulting formula is:
    \begin{equation}
        \sigma(^{0}(r-Y)) = \sqrt{\sigma_{m_{r}}^{2} + \sigma_{m_{Y}}^{2} + (z\sigma_{a_{1}})^{2}}.
        \label{eq:errors}
    \end{equation}
    We have dropped the $a_{1}^{2}\sigma_{z}^{2}$ term here, as $\sigma_{z}$ is negligible for galaxies with spectroscopic redshift measurements. 
    
    Thus, to calculate the uncertainty in $^{0}(r-Y)$  we will need to determine $\sigma_{a_{1}}$. For bands where $a_{1}$ is treated as linear in rest-frame colour, we determine $\sigma_{a_{1}}$ via bootstrap re-sampling \citep{efron1979}. We construct 100 sets of matched mean $^{0}(g-r)$ and $a_{1}$ values, each of equal size to the original data set used in the fit, but with each original (mean $^{0}(g-r)$, $a_{1}$) pair having equal probability of being chosen for each element of the new set of pairs that will be fit to (i.e., we perform bootstrapping of the original set of values from the fits in rest-frame colour bins, with replacement). For each of these hundred bootstrapped data sets we perform a Huber regression as in step (ii), fitting for \autoref{eq:a1_func}, which yields 100 sets of $b_{0}$ and $b_{1}$ values. Then for each galaxy we can determine 100 $a_{1}$'s via \autoref{eq:a1_func} by plugging in the galaxy's $^{0}(g-r)$. Finally, we calculate the standard deviation of the bootstrapped $a_{1}$ values for each object, providing an estimate of the appropriate $\sigma_{a_{1}}$ value for it.
    
    When $a_{1}$ is treated as a constant,  bootstrapping is not necessary to determine its uncertainty. Instead we estimate its value as $\sigma_{a_{1}} = \frac{\rm{NMAD}(a_{1})}{\sqrt{0.64N}}$, where $N$ is the number of colour bins used, as the standard deviation of the median of $N$ values drawn from a Gaussian distribution of standard deviation $\sigma$ is approximately $\frac{\sigma}{\sqrt{0.64N}}$ \citep{maindonald2010}.
    
    For convenience purposes, we generally quote photometric errors  (e.g., $\sigma_{m_{y}}$) separately from the uncertainty in the $\kcor$ ($z\sigma_{a_{1}}$). When a single net uncertainty is needed (e.g., for plots), we combine errors in quadrature, following \autoref{eq:errors}.
    
\end{enumerate}
\medskip
Tables of the $b_{0}$ and $b_{1}$ values neeeded to determine $a_{1}$ for arbitrary $^0(g-r)$ values, as well as the median $a_{1}$ values across all bins, are provided in \autoref{table:kcor_table_kcorrect} (\textsc{Kcorrect} derived) and \autoref{table:kcor_table_gswlc} (GSWLC-M2 derived) of \autoref{sec:appendix_tables}. The Python functions used to determine our $b_{0}$, $b_{1}$, and median $a_{1}$ values and related quantities in addition to sample code are provided at \href{https://github.com/cfielder/K-corrections}{our $K$-correction GitHub}\footnote{https://github.com/cfielder/K-corrections} page for public use.  These materials allow \kcors\ for low-redshift objects to be calculated easily for any of the bands considered in this paper.


\section{Results}
\label{sec:results}

We test our results by comparing to two other $\kcor$ methods performed on the same data: the standard software \textsc{Kcorrect},  
and SED fitting results from the GSWLC-M2 catalogue \citep{salim2016,salim2018} which are described in \autoref{subsection:data}. In our work the rest-frame values from which we determine our fits are from either the GSWLC-M2 catalogue calculations or the \textsc{Kcorrect} calculations. Depending on which \kcor\ method we wish to compare our calculations to we use the respective rest-frame colour. 

While there are discrepancies between \kcors\ calculated between any two methods, with our results in general agreement with other methods for well-behaved bands we can be confident in our simplified approach. 

\subsection{Distributions of Rest-frame Colours}
\label{subsection:rest-framecolor}

\begin{figure*}
    \centering
    \includegraphics[width=\linewidth]{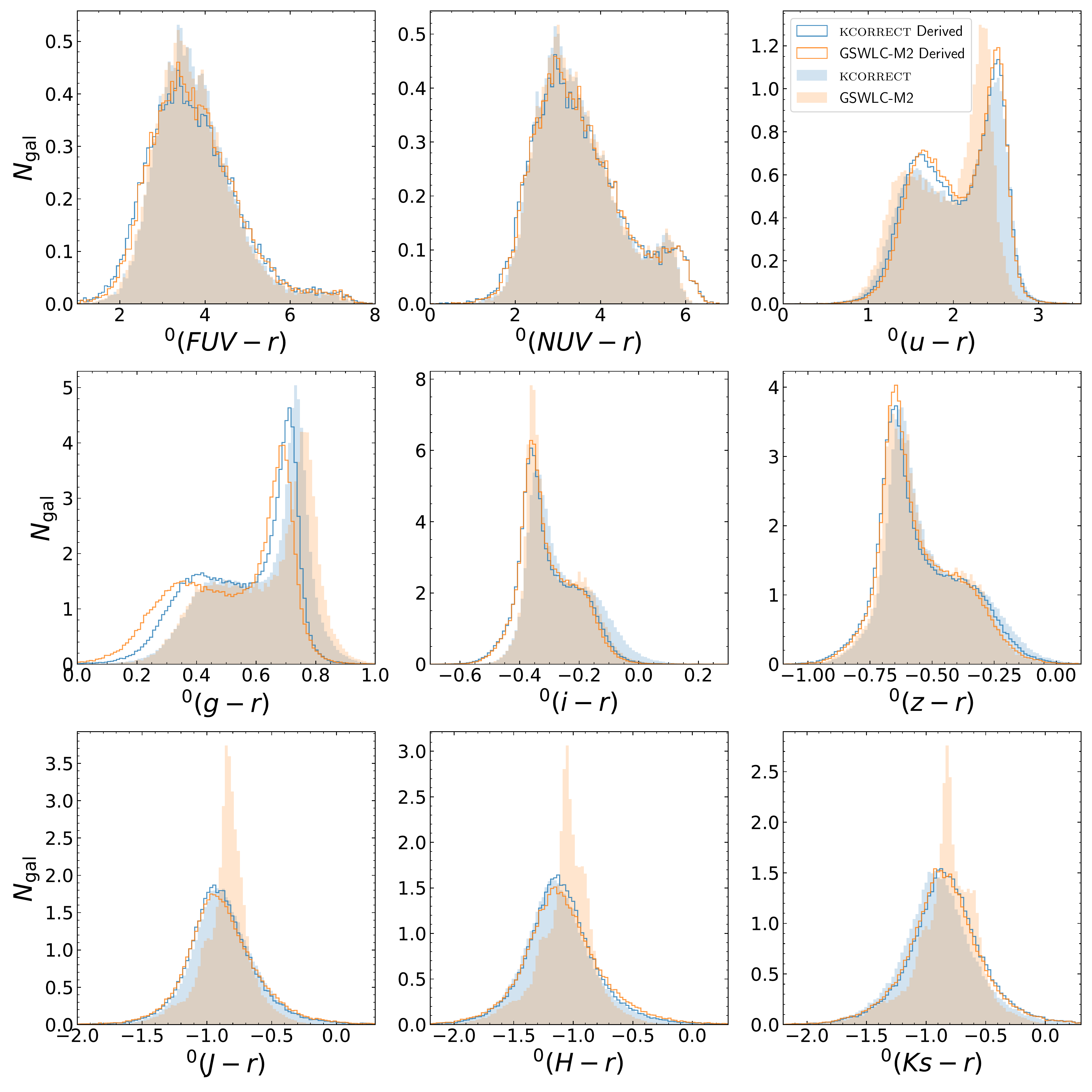}
    \caption{Histograms of rest-frame colours comparing \kcor\ methods for GALEX $FUV/NUV$, SDSS $ugiz$, and 2MASS $J/H/Ks$ bands. We show our calculated rest-frame colours derived from \textsc{Kcorrect} (blue line) and GSWLC-M2 (orange line) and those calculated using the \textsc{Kcorrect} software (blue shaded) or in the GSWLC-M2 catalogue (orange shaded). In general, our rest-frame colours derived from both \textsc{Kcorrect} and GSWLC-M2 are in excellent agreement with each other. In the 2MASS bands, our rest-frame colours derived from \textsc{Kcorrect} and GSWLC-M2 are self-consistent and in agreement with the \textsc{Kcorrect} colours; however, the GSWLC-M2 colours show more peaked distributions likely caused by uncertainties in modelling unusual older stellar populations (TP-AGB stars, post-AGB stars, and extreme HB stars). The offsets in $^{0}(g-r)$ are likely due to the fact that our rest-frame colours are derived from $^{0}(g-r)$ for both catalogues. Overall, our rest-frame colours for these bands are well-matched to each other and to those from \textsc{Kcorrect} and GSWLC-M2.
    }
    \label{fig:hists}
\end{figure*}

We start with a generalised comparison to \textsc{Kcorrect} and GSWLC-M2. In \autoref{fig:hists} normalised histograms depict results for colours across the wavelength range. We derive these respective to the catalogue to which they are being compared. For example we derive $^{0}(u-r)$ utilising $^{0}(g-r)$ from \textsc{Kcorrect} when comparing to \textsc{Kcorrect} results. The step histograms denote our derived results, and the lightly shaded histograms denote results from the two comparison methods. Blue histograms come from \textsc{Kcorrect} and orange histograms come from GSWLC. 

Foremost, we do not expect all results to be $1:1$ with each other due to the nature of fitting and how the \kcors\ are approximated in each method. The relatively large scatter of individual \kcors\ has been exemplified in the literature (see e.g., \citealt{omill2011}, Fig. 3, \citealt{chilingarian2010}, Fig. 2). We do find that, in general, the distributions of our $K$-corrected colours are in good agreement with those derived from other methods. In many cases our results are in better agreement with the respective comparison work than \textsc{Kcorrect} and GSWLC-M2 are with each other. However, there are some notable differences.  

\begin{figure*}
    \centering
    \includegraphics[width=\linewidth]{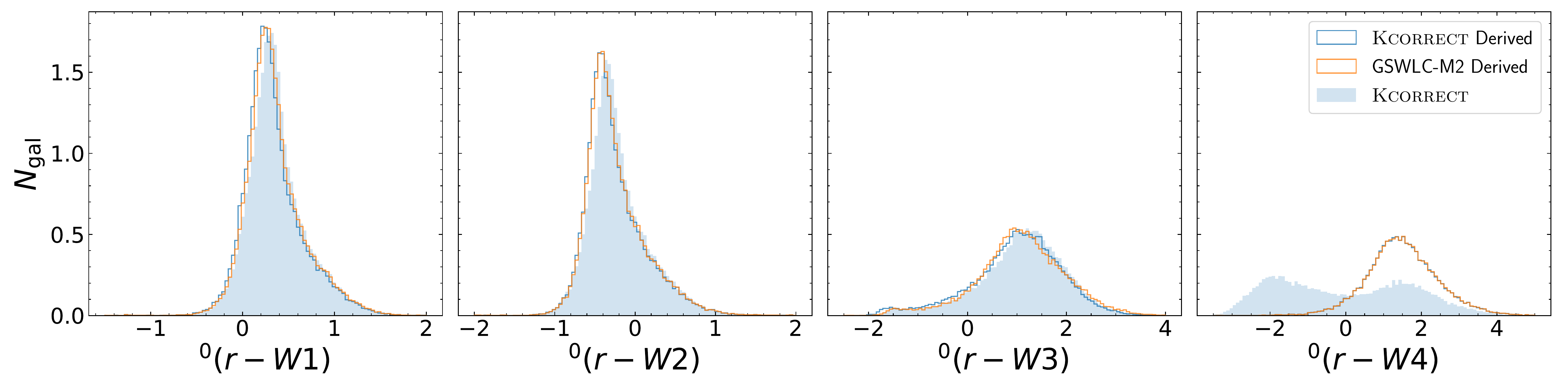}
    \caption{Same as \autoref{fig:hists} but for the WISE bands. GSWLC-M2 did not derive WISE colours, but we still derive rest-frame WISE colours using $^{0}(g-r)$ from both \textsc{Kcorrect} and GSWLC-M2. Our results derived from either source agree well with each other. Additionally, histograms match well with those of \textsc{Kcorrect}, except for the W4 band, which we suspect is a result of a template error in \textsc{Kcorrect}. This discrepancy in W4 highlights an advantage of deriving \kcors\ independent of templates in poorly constrained bands.}
    \label{fig:hists_wise}
\end{figure*}

Our results in the 2MASS bands from either derived source are in excellent agreement with each other and the results from \textsc{Kcorrect}. However, the GSWLC-M2 results are more peaked. Such discrepancies in the near-IR 2MASS bands in the GSWLC-M2 catalogue may result from a couple of factors. Templates in the IR are typically of lower quality than in the optical, particularly because there are few spectra fully observed from the UV-IR. Therefore some assumptions are required in the less constrained bands. For example, there is uncertainty of how asymptotic giant branch (AGB) stars contribute to stellar population synthesis models. In particular, thermally pulsating AGB (TP-AGB) stars may contribute a significant portion of light to the IR portion of an SED but how much is unknown \citep{conroy2013}. The IR portion of the SED is also further complicated by contributions of dust emission. It is likely the case here that the discrepancy of the GSWLC-M2 results is driven by the SED fits preferentially constrained in the UV and optical where models have more certainty.

While small, we note that discrepancies in the $FUV$, $NUV$, and $u$ bands are likely a result of (1) the high sensitivity of UV colours to minimal amounts of star formation, (2) the uncertainty of contributions of post-AGB and extreme horizontal branch stars, and (3) complexity of dust absorption and scattering \citep[see e.g.,][]{conroy2013}. Simple stellar populations, which are the backbone of SPS model SED based approaches, may not adequately capture the UV sensitivity to star formation, and dust and unusual stellar populations further complicate these effects.

In \autoref{fig:hists} it appears that there are also discrepancies of note in the optical bands (particularly $g$ and $i$). However, across wavelength space the differences between results are no more than $\sim 0.1$ magnitudes and agree on average at the $0.05$ magnitude level after exploration of the NMAD and root mean squared error between our results and the respective literature result. Differences are more pronounced in the optical given the narrow range of colour space that they span. While the $^{0}(g-r)$ results presented here are calculated in bins of $^{0}(g-r)$, which is redundant, we also derived $^{0}(g-r)$ from computing $^{0}(g-i)$ and $^{0}(r-i)$ in bins of $^{0}(g-r)$ and then combining results. The results were similar, so we present the more straightforward result here. We want to emphasise that the bluest galaxies in \textsc{Kcorrect} and GSWLC-M2 have derived $^{0}(g-r)$ colours redder than ours starting at $z\sim0.03$. Additionally, the rest-frame colour derived from the SED fitting techniques is \textit{redder} than the observed colour for these very blue objects, contrary to the expected blue-ward shift in the optical post \kcor . The likely culprit is either a lack of SED templates for the most extremely blue galaxies, or some lacking or excess contribution from the dust absorption and nebular emission models. Investigations in small redshift bins of rest-frame $(g-r)$ colour show that the red sequence is more consistent at different redshift with our \kcor\ method than $^{0}(g-r)$ colour derived using \textsc{Kcorrect}.

\autoref{fig:hists_wise} depicts the same comparisons as \autoref{fig:hists} for the WISE bands. The GSWLC-M2 did not perform SED fitting directly with the WISE bands \citep{salim2018}, so those values are not included. Overall, we observe very consistent results with those of \textsc{Kcorrect} except in the $W4$ band. This miss-match in distributions between results of \textsc{Kcorrect} and the distributions of the observed photometry and our own derived rest-frame photometry can largely be attributed to the lack of dust emission modelling in \textsc{Kcorrect}. Therefore any rest-frame $W2$ through $W4$ results obtained via \textsc{Kcorrect} could be incorrect due to the lack of these types of models.

\begin{figure}
    \centering
    \includegraphics[width=0.96\linewidth]{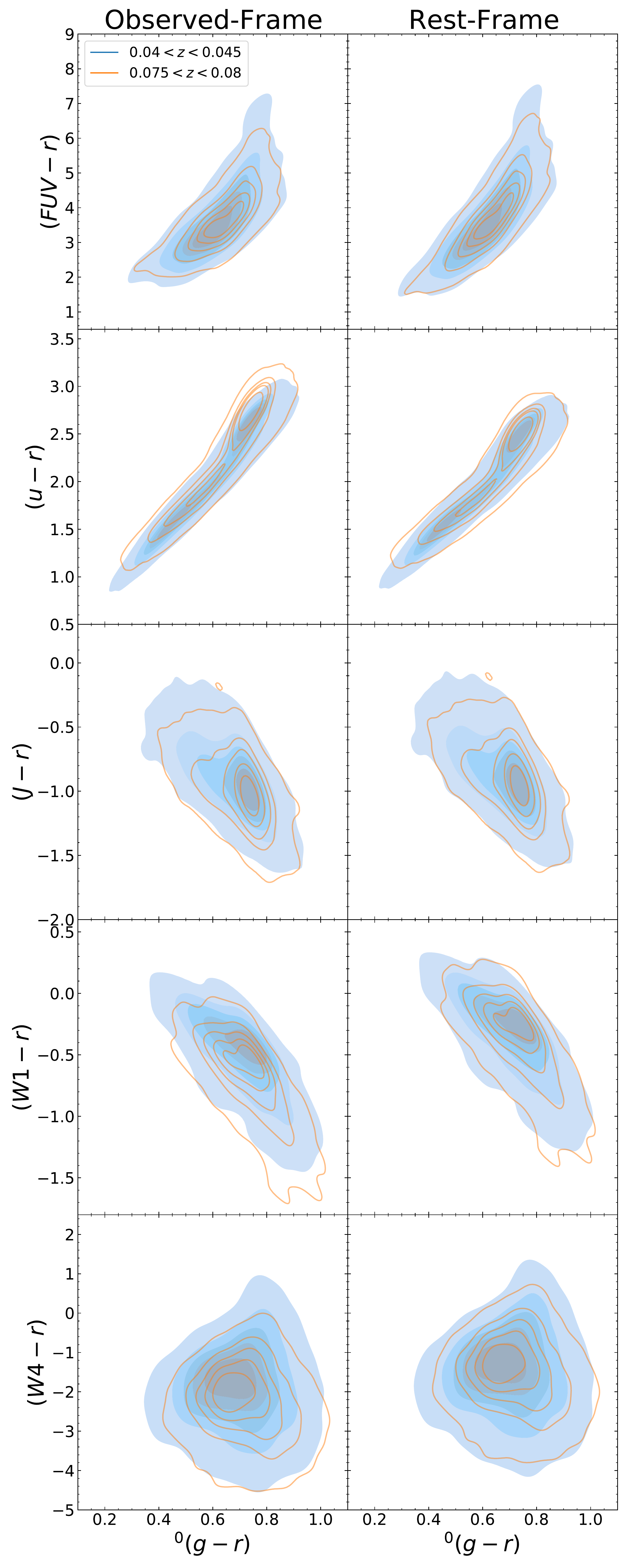}
    \caption{Contour plots of galaxies in two narrow redshift bins $0.04 < z < 0.045$ (blue) and $0.075 < z < 0.08$ (orange). We plot a variety of colours across our wavelength range as a function of $^{0}(g-r)$. The left column shows these colours in the observed-frame and the right shows these colours in the rest-frame, using our \kcor\ results. We plot narrow redshift bins as the observed SEDs of the galaxies should be approximately the same. It is evident that for all colours the contours of both redshift bins align closely in the rest-frame in contrast to the observed-frame. This is the expected result for successful \kcors\ and similar results are observed in \textsc{Kcorrect} and GSWLC-M2 (save for $(W4-r)$).}
    \label{fig:contours}
\end{figure}

To check the performance of our \kcors\ we examine colours in small redshift bins. We choose narrow bins as there should be minimal colour variance across the galaxy sample, meaning they should all have approximately the same observed frame SED. \autoref{fig:contours} shows results of this analysis for a selection of colours across our wavelength range. We plot contours for $0.04 < z < 0.045$ (blue) and $0.075 < z < 0.08$ (orange). On our x-axis we plot $^{0}(g-r)$. The left column depicts contours using observed colours and the right column depicts contours for rest-frame colours. These specific rest-frame colours are derived from \textsc{Kcorrect} $^{0}(g-r)$. For objects that have been properly $K$-corrected, the contours should line up better in the rest-frame column than the observed-frame column. This is indeed the case for our bands, which indicates that our \kcors\ are working as expected. Plots made for \textsc{Kcorrect} and GSWLC-M2 rest-frame colours can be made, with similar results, save for $(W4-r)$ which is dramatically offset in the \textsc{Kcorrect} results. While $(FUV-r)$ exhibits minimal change, this is also the case for \textsc{Kcorrect} and GSWLC-M2.

\subsection{Extrapolations to z=0}
\label{subsection:extrapolation}

\begin{figure}
    \centering
    \includegraphics[width=\linewidth]{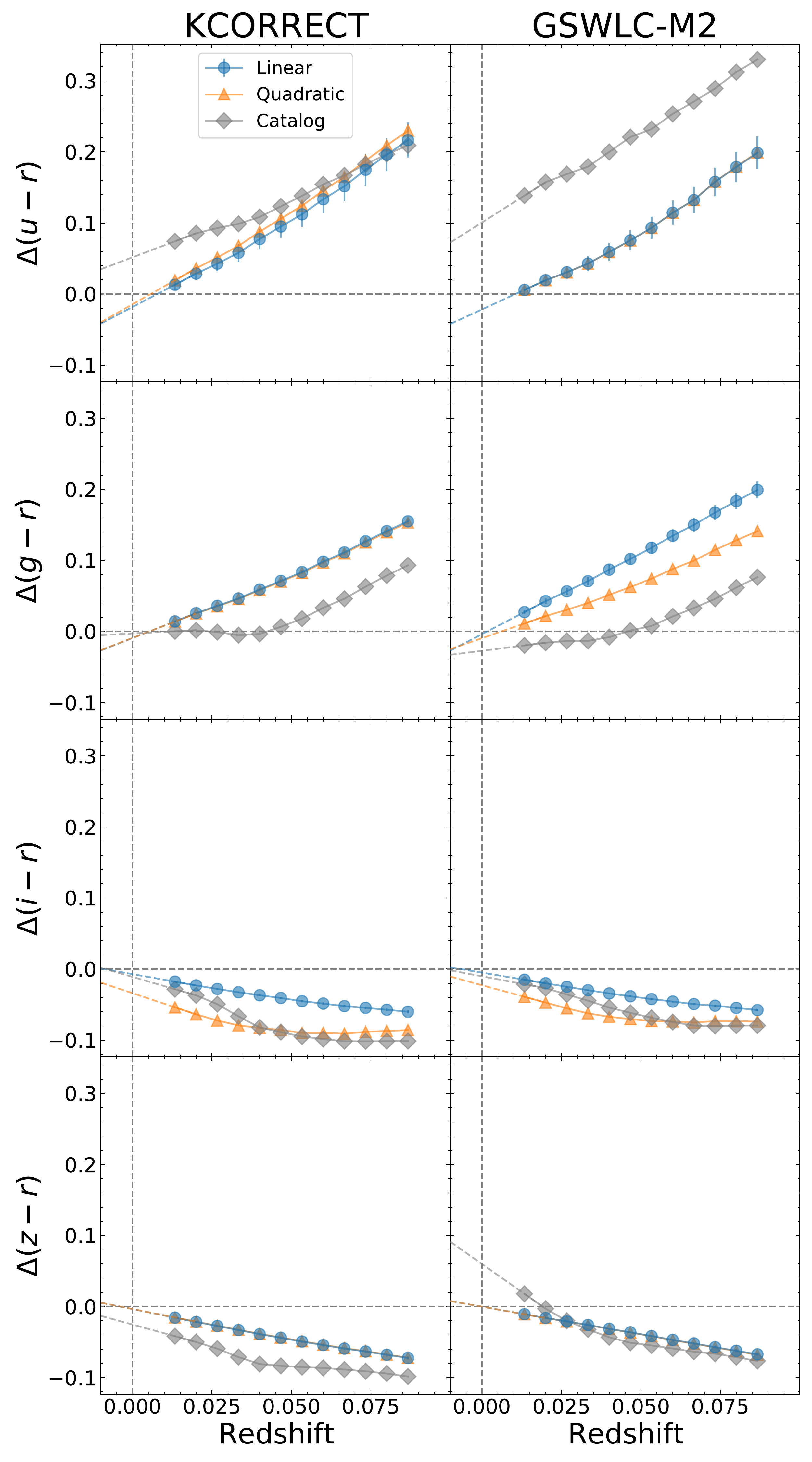}
    \caption{Plots of the mean difference between observed and rest-frame colour (e.g., $\Delta(u-r) = (u-r) - ^{0}(u-r)$) as a function of 12 redshift bins for optical colours. Blue points depict our $K$-corrected results with a linear initial fit, and orange points depict results for a quadratic fit. Grey points show results from the corresponding source denoted at the column title. The coloured dashed lines depict a linear extrapolation to redshift $0$. Our results all approach the origin closely, as they should by construction and under the expectation that are redshift $0$ the \kcor\ should be $0$. However, this is not necessarily the case for \kcors\ that rely on SED fitting.}
    \label{fig:delta_optical}
\end{figure}

\begin{figure}
    \centering
    \includegraphics[width=\linewidth]{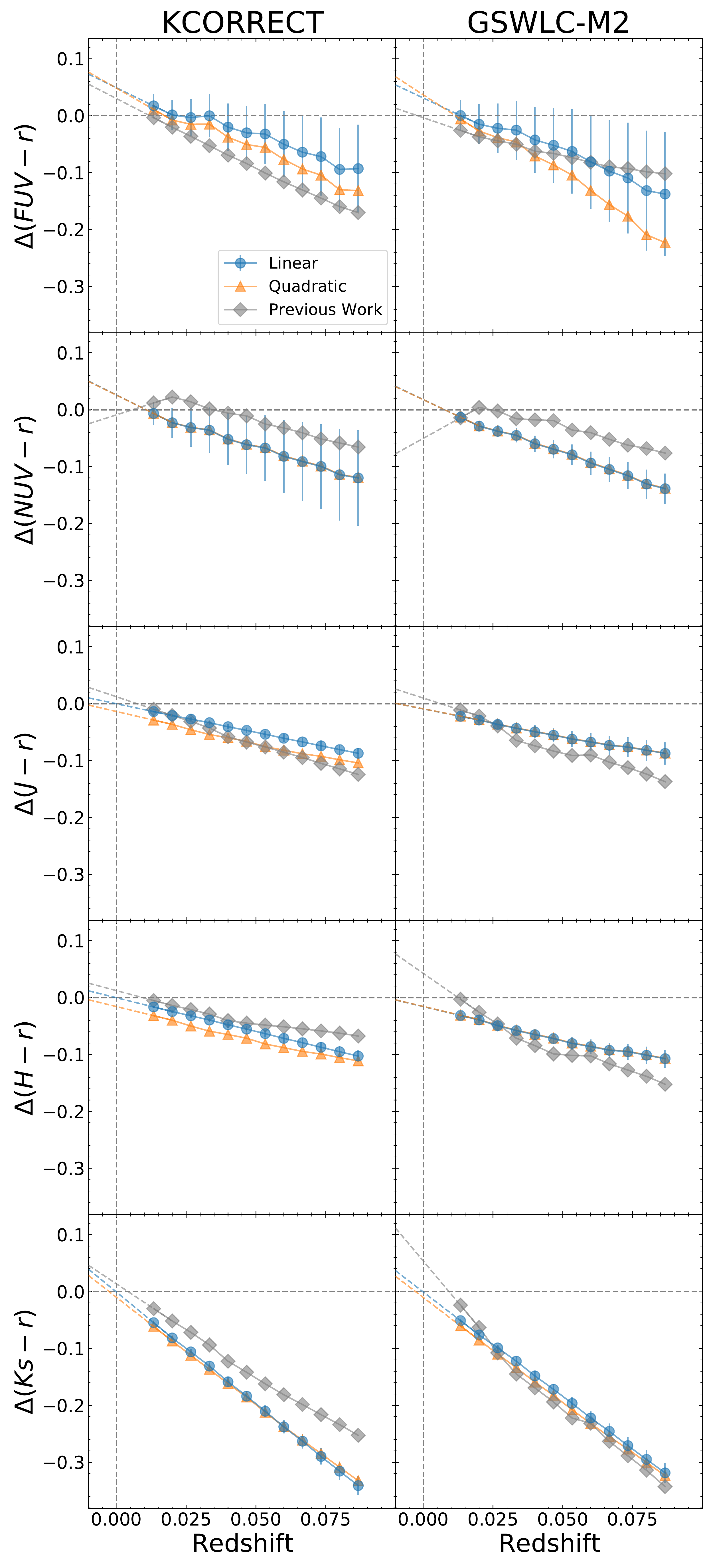}
    \caption{Same as \autoref{fig:delta_optical} but for UV and near-IR bands. Our results perform just as well as if not better than those from SED fitting.}
    \label{fig:delta_nonoptical}
\end{figure}

At redshift $0$, by definition a \kcor\;must be $0$. This means that any difference between observed colours (or magnitudes) and rest-frame colours (or magnitudes) at redshift $0$ must also be $0$. As we showed in \autoref{subsection:calculation} our \kcors\ are defined to follow this relationship. However, this is not enforceable for methods that incorporate SED/template fitting. We compare our $\kcor$ colours to that of \textsc{Kcorrect} and GSWLC-M2 in \autoref{fig:delta_optical} and \autoref{fig:delta_nonoptical} for optical and UV/near-IR bands, respectively. We plot the colour difference as a function of redshift where, for example, $\Delta(u-r) = (u-r) - ^{0}(u-r)$ or the difference between the observed-frame colour and rest-frame colour. We specifically plot the means of these colour differences across 12 bins in redshift space. Blue points correspond to $K$-corrected results using our method with a linear fit and orange points correspond to $K$-corrected results using our method with a quadratic fit. Grey points correspond to results from \textsc{Kcorrect} (left column) and GSWLC (right column).

The blue points also have representative error bars plotted that include both systematic errors and $\kcor$ errors, for which the $\kcor$ errors are described in (v) of \autoref{subsection:calculation}. We must take a few additional steps for these plots, as we are showing the means in bins. The error attributed to the $\kcor$ does \textit{not} diminish when determining means. So for the $\kcor$ error we simply compute $\langle\sigma_{a_{1}}z\rangle$. We define the random error on the mean as $\sigma_{sys} = \frac{\sigma(\Delta)}{\sqrt{N_{bin}}}$. The total error on the means is thus these two contributions added in quadrature. 

Because our galaxy sample only extends down to redshift $0.01$, we must extrapolate to $z=0$. This is done with a linear interpolation for colour difference as a function of the redshift of the bin centre ($z_{\rm{cen}}$), utilising the $\tt{scipy.interpolate.interp1d}$ class from \citet{SciPy2020}. These extrapolations are plotted as dashed lines, which ideally should pass through the origin. Our linear and quadratic fits do so in most cases (with some scatter), in contrast to fits from \textsc{Kcorrect} and GSWLC-M2 for many bands. This directly demonstrates the improved accuracy of our $\kcor$ method compared to SED-fitting $\kcor$ methods. Additionally these figures exemplify that in most cases a linear function of redshift is a better choice than a quadratic one for deriving our relationship between observed colour, redshift, and rest-frame colour. First, the linear fit derives very comparable results to the quadratic fit and is much simpler to implement. Second, the linear fit results more consistently pass through the origin than the quadratic fit results. It is also worth noting that despite the difference in the techniques, the relative shapes of the results in both \textsc{Kcorrect} and GSWLC-M2 agree well. Likewise the relative shapes of our results, regardless for which source the rest-frame band is used, are also in agreement in both linear and quadratic initial fits.

\subsection{Spectral Energy Distributions}
\label{subsection:SEDs}

Spectral energy distributions (SEDs) are an excellent test of our $\kcor$ results. In this subsection we compare our \kcored\ rest-frame photometry to that of our other sources of \kcored\ photometry and the observed photometry for a couple of galaxies. Because our galaxy sample is at such a low redshift, the $\kcor$ in each band should be relatively small and mirror the shape of the observed SED.

To determine luminosity we use similar derivations to those presented in Section 4.3 of \citet{fielder2021}, which we also summarise here. Because the $r$-band serves as the anchor in our $\kcor$ derivations we can derive fluxes relative to the $r$-band. We use $^{0}M_{r}$ from \textsc{Kcorrect} and GSWLC-M2 respective to which $\kcor$ source we compare to. The luminosity of $r$-band is derived as 
\begin{equation}
        \log{(\nu_{r}L_{\nu,r})} = \log{\nu_{r}} + \frac{(^{0}M_{r} - 34.04)}{-2.5},
\label{eq:lum_r}
\end{equation}
which is derived from the relation for converting flux to AB magnitudes in combination with the area of a 10 pc radius sphere to convert flux to luminosity. The error on this quantity via propagation of errors is equivalent to 
\begin{equation}
    \sigma_{\log(\nu_{r}L_{\nu,r})} = 0.4\sigma_{^{0}M_{r}}.
\end{equation}

For the other bands we use our derived colours $^{0}(r-Y)$ to first determine the flux ratio relative to the $r$-band
\begin{equation}
        \log{\left( \frac{f_{\nu,r}}{f_{\nu,Y}}\right)} = \frac{^{0}(r-Y)}{-2.5}.
\label{eq:fracflux}
\end{equation}
This formula is based on the definition of observed AB magnitude. We can then derive luminosity for other bands utilising the equivalence of flux ratios to luminosity ratios. Our final SED luminosities are
\begin{equation}
        \log{(\nu_{Y}L_{\nu,Y})} = \log{(L_{\nu,r})} - \log{\left( \frac{f_{\nu,r}}{f_{\nu,Y}}\right)} + \log{(\nu_{Y})},
\label{eq:lum}
\end{equation} 
where $\log{(\nu_{Y}L_{\nu,Y})}$ is in units of Watts. The corresponding errors are defined as 
\begin{equation}
    \sigma_{\log{\nu L_{\nu,x}}} = 0.4\sqrt{\sigma_{m_{r}}^{2} + \sigma_{m_{Y}}^{2} + (z\sigma_{a_{1}})^{2} \sigma_{^{0}M_{r}}^{2}},
\label{eq:sed_err}
\end{equation}
which can be derived from propagation of errors and plugging \autoref{eq:rest_color} into \autoref{eq:fracflux}.

\begin{figure*}
    \centering
    \includegraphics[width=\linewidth]{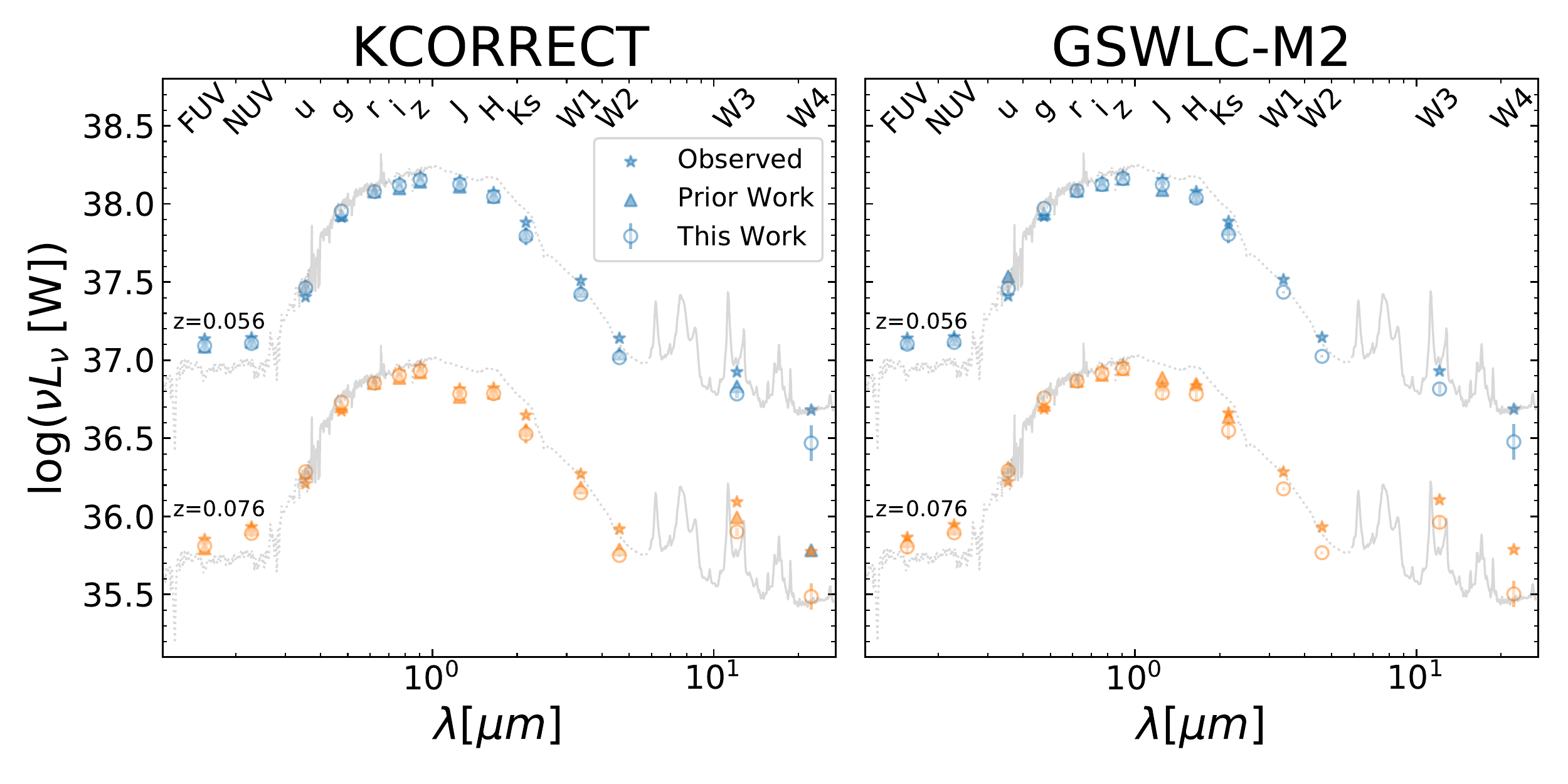}
    \caption{Derived SEDs for two galaxies from our sample which share the same $^{0}(g-r) = 0.667$ but reside at two different redshifts. The lower redshift galaxy is plotted with blue points and offset from the higher redshift galaxy in the y-axis for clarity. Stars are plotted at the observed fluxes. Triangles are plotted at the luminosity determined from the rest-frame magnitudes derived via \textsc{Kcorrect} (left panel) and GSWLC-M2 (right panel). Circles with error bars show our resulting luminosities from our $\kcor$ method. As a means of comparison we plot the spectrum of NGC 4138 from the \citet{brown2014} galaxy SED atlas in grey, with dashed lines showing modelled portions of the spectra and solid lines showing the observed portions of the spectra. This template had the lowest $\chi^{2}$ of the difference between the photometry of the two galaxies shown and the template (normalised in $r$-band). While we do not expect a direct match to a template, we provide it as a means of comparison to help guide the eye. We can conclude that our $\kcor$ method provides reasonable estimates for even the weakly constrained WISE bands, where \textsc{Kcorrect} fails in $W4$ and GSWLC lacks. The rest-frame $W4$ prediction of the higher redshift galaxy lies below the $y$-axis range and the rest-frame prediction for the lower redshift galaxy overlaps with the observed photometry of the higher redshift galaxy. Given the redshifts of these objects and our data driven approach, our \kcored\ magnitudes cannot significantly deviate from the observed photometry, which allows us to $K$-correct even low signal-to-noise bands like $W3$ and $W4$.
    }
    \label{fig:sed}
\end{figure*}

In \autoref{fig:sed}, we plot the SEDs of two galaxies from our sample. These galaxies share the same $^{0}(g-r)$ colour of $0.667$ but reside at two different redshifts with the lower redshift galaxy plotted in blue and the higher redshift galaxy plotted in orange. The normalisation of the lower redshift object is offset such that the two galaxies can be plotted without overlap. Open circles with error-bars show results from our work, which are derived using either a linear or constant $a_{1}$ in \autoref{subsection:calculation}. For all bands other than $r$, we do not plot the $\sigma_{^{0}M_{r}}$ term in \autoref{eq:sed_err} as this error is covariant across all bands. This quantity is small and challenging to infer from the plots; $\sigma_{^{0}M_{r}} = 0.0012$ for the low redshift plot and $\sigma_{^{0}M_{r}} = 0.0032$ for the higher redshift plot. Triangular points show the \kcors\ for these galaxies from \textsc{Kcorrect} (left panel) and GSWLC-M2 (right panel). Lastly, star points show the observed fluxes of the same galaxies, normalised to match in the $r$-band.

As a means to compare these two galaxies we also plot an SED template galaxy from the \citet{brown2014} nearby galaxy SED atlas - NGC 4138. The magnitudes of the templates were converted to fluxes (as described in 4.2.2 of \citealt{fielder2021} utilising the AB magnitude definition), and then normalised in $r$-band. For each of the 129 galaxy SED templates in the atlas we calculate the chi-squared of the difference between the brightness of each of our selected galaxies and that of the templates:
\begin{equation}
    \chi^{2} = \sum \left( \frac{\log{\nu L_{\nu}^{\rm{atlas}}}-\log{\nu L_{\nu}^{\rm{galaxy}}}}{\sigma_{\log{\nu L_{\nu}}}} \right)^{2}.
\end{equation}
\label{eq:chisq}
$\sigma_{\log{\nu L_{\nu}}}$ combines in quadrature the total error in the galaxy SED for a given band, the uncertainties in the \citet{brown2014} photometry, and $\log_{10}$ (1.1), which corresponds to a 10\% error in $\nu L_{\nu}$. This additional 10\% error is added to account for systematic uncertainties in the photometry such that optical bands do no dominate the $\chi^{2}$ values given their intrinsically small uncertainties. 

The template plotted in grey in \autoref{fig:sed} corresponds to the spectrum of the NGC 4138 template, which had the smallest $\chi^{2}$ compared to \textit{both} galaxies selected from our sample. Dashed portions of the plotted template indicate regions of the spectrum that were modelled, while solid portions of the template indicate regions of the spectrum that were observed. We do not expect this template to directly match the photometry of either selected galaxy but to provide a means of comparison. This spectrum also exemplifies the use of modelling required in the UV and IR for empirical SED templates.

One can draw several conclusions from \autoref{fig:sed}. First, our \kcored\ photometry is reasonably close to that of other methods of producing \kcored\ photometry in addition to the observed photometry. At our low redshift range the rest-frame photometry should not be substantially different from the observed photometry, which is the case for our results. When combined, we have results that resemble an SED reasonably well, while making minimal assumptions about how an SED \textit{should} look.

Second, it is evident that \textsc{k-correct} fails in giving a reasonable result for W4 band \kcors\, which is further supported by the results shown in \autoref{fig:hists_wise}. In \autoref{fig:sed} the brightness of the lower redshift galaxy's W4 band from \textsc{Kcorrect} is so low that it overlaps with the other galaxy's observed brightness. For the higher redshift galaxy the W4 point is below the $y$-axis range. Our \kcor\ approach supplies reasonable results for $W4$. By construction our method cannot stray too far from the observed photometry, so even bands like $W4$ that have low signal-to-noise ratios can be reasonably constrained. In that same vein our \kcor\ method has more flexibility than standard SED fitting where multiple bands are forced to match the templates.

\section{Conclusions}
\label{sec:conclusions}

In this work, we present a new, empirically-driven method for obtaining \kcors. Our method is inspired by the polynomial based $\kcor$ methods of \citet{beare2014} and \citet{chilingarian2010}, where we exploit the parameterisation of \kcors\ by an analytical function of redshift and colour \citep[see also][]{willmer2006,omill2011}. At low redshift ($z < 1$), SEDs fall into a single parameter family where a single rest-frame colour allows us to infer the full SED shape - an inference made from results describing spectra by a single parameter in reduced dimensional space \citep{connolly1995,madgwick2003a,madgwick2003b}. Our method limits the dependence on SED templates by interpolating using a rest-frame colour for which SEDs are well-constrained. We perform a series of linear fits to construct a function that maps observed colour and redshift to rest-frame colour.

There are a number of pieces of evidence that prove our data driven approach to determining \kcors\ is yielding sensible results. Throughout this work we compare to the  oft-used \textsc{Kcorrect v4.3} software of \citet{blanton2007} and the derived results in the GSWLC-M2 catalogue \citep{salim2016,salim2018}, both of which utilise SED fitting in different ways (see \autoref{subsection:kcorrection_catalogues}). Our \kcored\ photometry is in agreement with \kcored\ photometry of the same objects for both comparison methods, particularly at UV and optical wavelengths (see e.g., \autoref{fig:hists}). Likewise our \kcored\ photometry produces sensible looking galaxy SEDs, similar to both comparison methods (see \autoref{fig:sed}). We also have better agreement across redshift in colour-colour space with our rest-frame colours than our observed-frame colours, as exemplified in \autoref{fig:contours}. This is the expected behaviour of photometry that has been correctly \kcored. Lastly, our \kcors\ go to zero at redshift zero, as shown in \autoref{fig:delta_optical} and \autoref{fig:delta_nonoptical}, as expected by the definition of the \kcor. However, SED fitting methods cannot directly enforce this, as exemplified by their differences in the figures. Our method is driven by and stays close to the observations in the band of interest rather than follow a model which can yield larger deviations. This allows us to $K$-correct bands that are poorly constrained by templates and have low signal-to-noise ratios.

There are also notable discrepancies between our \kcored\ photometry and that of \textsc{Kcorrect} and GSWLC-M2, which mostly occur in the infrared bands. First, as seen in \autoref{fig:hists}, the rest-frame photometry in the 2MASS $JHKs$ bands is more peaked compared to the other histograms. However, our derivations from both \textsc{Kcorrect} and GSWLC-M2 are in excellent agreement with each other and that of \textsc{Kcorrect}. This highlights one of advantage of our approach. SED model uncertainties in the IR due to contributions from AGB stars may result in incorrect \kcors. Higher signal-to-noise optical portions may also drive SED fits resulting in incorrect \kcors. Empirical SED models are also more poorly constrained at non-optical wavelengths due to a lack of spectral measurements that in turn produce uncertain SED fits. The other notable discrepancy is that of the WISE $W4$ band, noted in \autoref{fig:hists_wise} and \autoref{fig:sed}. The predicted \textsc{Kcorrect} $W4$ band rest-frame photometry is severely offset from our predictions, while our results utilising either \textsc{Kcorrect} or GSWLC-M2 optical colours are in good agreement with each other and with \textsc{Kcorrect} results in the other WISE bands. We suspect that the \textsc{Kcorrect} templates are making un-physical assumptions, yielding an unexpected result. Most notable is the lack of dust emission modelling in \textsc{Kcorrect}, a feature of increasing importance for the WISE bands. In contrast, our data-driven approach entirely avoids issues of template fit failures and model issues by only relying on well-constrained rest-frame optical colours.   

Our data driven \kcor\ approach is particularly suitable for calculations in bands where templates are not well constrained or readily available. In our previous work of \citet{fielder2021} in constructing a UV-IR SED for the Milky Way we required WISE-band \kcors\ - bands sensitive to dust content and dust luminosity (which are consequently a constraint on star formation rate, see e.g., \citealt{salim2018}). Because WISE-band \kcors\ were not readily accessible we developed the approach presented here for determining \kcors\ . 


\section*{Acknowledgements}

Authors Catherine Fielder, Brett Andrews, and Jeff Newman gratefully acknowledge support from NASA Astrophysics Data Analysis Program grant number 80NSSC19K0588 which made this research possible.

This work made use of \textsc{Python}, along with many community-developed or maintained software packages, including IPython \citep{ipython}, Jupyter (\http{jupyter.org}), Matplotlib \citep{matplotlib}, NumPy \citep{numpy}, Pandas \citep{pandas}, scikit-learn \citep{scikit-learn}, and SciPy \citep{SciPy2020}.
This research made use of NASA's Astrophysics Data System for bibliographic information.

This paper was based in part on observations made with the NASA Galaxy Evolution Explorer. GALEX is operated for NASA by the California Institute of Technology under NASA contract NAS5-98034.

This publication makes use of data products from SDSS-IV. Funding for the Sloan Digital Sky 
Survey IV has been provided by the 
Alfred P. Sloan Foundation, the U.S. 
Department of Energy Office of 
Science, and the Participating 
Institutions. 

SDSS-IV acknowledges support and 
resources from the Center for High 
Performance Computing  at the 
University of Utah. The SDSS 
website is www.sdss.org.

SDSS-IV is managed by the 
Astrophysical Research Consortium 
for the Participating Institutions 
of the SDSS Collaboration including 
the Brazilian Participation Group, 
the Carnegie Institution for Science, 
Carnegie Mellon University, Center for 
Astrophysics | Harvard \& 
Smithsonian, the Chilean Participation 
Group, the French Participation Group, 
Instituto de Astrof\'isica de 
Canarias, The Johns Hopkins 
University, Kavli Institute for the 
Physics and Mathematics of the 
Universe (IPMU) / University of 
Tokyo, the Korean Participation Group, 
Lawrence Berkeley National Laboratory, 
Leibniz Institut f\"ur Astrophysik 
Potsdam (AIP),  Max-Planck-Institut 
f\"ur Astronomie (MPIA Heidelberg), 
Max-Planck-Institut f\"ur 
Astrophysik (MPA Garching), 
Max-Planck-Institut f\"ur 
Extraterrestrische Physik (MPE), 
National Astronomical Observatories of 
China, New Mexico State University, 
New York University, University of 
Notre Dame, Observat\'ario 
Nacional / MCTI, The Ohio State 
University, Pennsylvania State 
University, Shanghai 
Astronomical Observatory, United 
Kingdom Participation Group, 
Universidad Nacional Aut\'onoma 
de M\'exico, University of Arizona, 
University of Colorado Boulder, 
University of Oxford, University of 
Portsmouth, University of Utah, 
University of Virginia, University 
of Washington, University of 
Wisconsin, Vanderbilt University, 
and Yale University.

This publication makes use of data products from the Two Micron All Sky Survey, which is a joint project of the University of Massachusetts and the Infrared Processing and Analysis Center/California Institute of Technology, funded by the National Aeronautics and Space Administration and the National Science Foundation.

This publication makes use of data products from the Wide-field Infrared Survey Explorer, which is a joint project of the University of California, Los Angeles, and the Jet Propulsion Laboratory/California Institute of Technology, funded by the National Aeronautics and Space Administration.

\section*{Data Availability}

Data used in this article is provided publicly at  \href{https://salims.pages.iu.edu/gswlc/}{the GSWLC website.} The reduced data used in this article is provided publicly on \href{https://github.com/cfielder/K-corrections/tree/main/Catalogs}{our project GitHub}. The ReadMe also provides a detailed description of the data products. Additional data is available upon request.



\bibliographystyle{mnras}
\bibliography{refs} 


\clearpage
\appendix

\section{Tables of Polynomial Coefficients}
\label{sec:appendix_tables}

In this section we present the derived intercept ($b_{0}$) and slope ($b_{1}$) for the secondary fits described in step (ii) of \autoref{subsection:calculation}; these values are sufficient to enable the \kcor\ to be calculated for any object at low $z$ given its rest-frame $g-r$ colour and redshift. For bands where a constant $a_{1}$ is favoured by the information criterion (cf. step (iii) of \autoref{subsection:calculation}),  we present the median $a_{1}$ value instead (``med. $a_{1}$''). \autoref{table:kcor_table_kcorrect} presents results calculated using \textsc{Kcorrect} $^{0}(g-r)$ serves as the reference rest-frame colour, while \autoref{table:kcor_table_gswlc} presents results where GSWLC-M2 $^{0}(g-r)$ is used instead.  These tables all use the optical SDSS $r$ band as an anchor band; i.e., the \kcors\ tabulated are those needed to convert observed $r-Y$ colour, where $Y$ is the band listed in a given row, to rest-frame $r-Y$.

The $b_{0}$ and $b_{1}$ values in these tables specifically are determined from \autoref{eq:a1_func}. The reader can use these quantities to determine $a_{1}$ for any object given its rest-frame colour. Then by applying \autoref{eq:polynomial2} one can obtain the rest-frame $(r-Y)$ colour for any band $Y$ of interest. 

For example, if one wished to determine rest-frame quantities for the optical $z$-band from these tables, one can determine the $a_{1}$ value for an object by applying the relation $a_{1} = b_{0} + b_{1}^{0}(g-r)$, where $^{0}(g-r)$ is the rest-frame colour of the chosen galaxy. Then one can obtain its for rest-frame colour by calculating $^{0}(r-z) = (r-z)_{\rm{obs}}-a_{1}z$. For bands that use median $a_{1}$, the first equation can be bypassed and the median $a_{1}$ value used instead. Hence, to determine rest-frame ($r-W4$) colours one can simply compute $^{0}(r-W4) = (r-W4)_{\rm{obs}}-($med. $a_{1})z$.  If one desires to determine an absolute magnitude rather than a rest-frame colour, this can be determined by next applying \autoref{eq:absmag}.

In addition to these tables we provide code for determining med. $a_{1}$, $b_{0}$ and $b_{1}$ or other related quantities at \href{https://github.com/cfielder/K-corrections}{our GitHub repository}. Full public access is provided for adaptation to any other project, under a CC BY-SA 4.0 license. 

\begin{table}
\centering
\begin{tabular}{|r||l|l|l|}
\hline
 \multicolumn{4}{|c|}{\textsc{Kcorrect} Derived Coefficients} \\
\hline
\hline
Passband & med. $a_{1}$ & $b_{0}$ & $b_{1}$ \\
\hline
FUV & & -18.557 & 30.780 \\
NUV & & -7.311 & 13.429 \\
u &  & 2.503 & -7.553 \\
g &  & 0.172 & -2.849 \\
i &  & 2.609 & -2.887 \\
z &  & 1.816 & -1.440 \\
J & 1.009 &  &  \\
H & 1.189 &  &  \\
Ks & 3.946 &  &  \\
W1 & 3.945 &  &  \\
W2 & 5.520 &  &  \\
W3 & 6.252 &  &  \\
W4 & 9.394 &  &  \\
\hline
\end{tabular}
\caption{Table of \textsc{Kcorrect} $^{0}(g-r)$ derived quantities for the median $a_{1}$'s, intercept ($b_{0}$), and slope ($b_{1}$) of the secondary fits for all bands considered in this work. In some cases, information criteria favour simply using the median derived $a_{1}$ values, while for others it is preferred to apply the linear relation $a_1 = b_{0} + b_{1}^{0}(g-r)$. In either case, rest-fame colours can then be determined via the relation $^{0}(r-Y) = (r-Y)_{\rm{obs}}-a_{1}z$, where $Y$ is the band for which rest-frame quantities are to be derived.}
\label{table:kcor_table_kcorrect}
\end{table}

\begin{table}
\centering
\begin{tabular}{|r||l|l|l|}
\hline
 \multicolumn{4}{|c|}{GSWLC-M2 Derived Coefficients} \\
\hline
\hline
Passband & med. $a_{1}$ & $b_{0}$ & $b_{1}$ \\
\hline
FUV & & -11.243 & 19.597 \\
NUV & & -4.251 & 8.790 \\
u &  & 3.986 & -9.244 \\
g &  & 1.089 & -4.069 \\
i &  & 2.088 & -2.089 \\
z & 0.777 &  &  \\
J & & 4.345 & -4.506 \\
H & & 6.955 & -7.721 \\
Ks & 3.687 &  &  \\
W1 & 3.548 &  &  \\
W2 & 5.391 &  &  \\
W3 & & 21.736 & -25.185 \\
W4 & 9.277 &  &  \\
\hline
\end{tabular}
\caption{As \autoref{table:kcor_table_kcorrect}, but for quantities derived utilising GSWLC-M2 $^{0}(g-r)$. Note that the set of passbands for which using the median $a_{1}$ value was favoured differed in this case.}
\label{table:kcor_table_gswlc}
\end{table}


\bsp	
\label{lastpage}
\end{document}